\pdfoutput=1

\documentclass[12pt,a4paper]{article}

\usepackage{ifthen} 

\usepackage{amsfonts}
\usepackage{makecell}

\newboolean{pdflatex}
\setboolean{pdflatex}{true} 

\newboolean{articletitles}
\setboolean{articletitles}{true} 

\newboolean{uprightparticles}
\setboolean{uprightparticles}{false} 

\def\paperauthors{LHCb collaboration} 
\def\paperasciititle{Search for CP violation in Ds+ -> K-K+K+ and D+ -> K-K+K+ decays} 
\def\papertitle{Search for  \CP violation in $\Dspp\to \Km\Kp\Kp$ decays} 
\def\paperkeywords{{High Energy Physics}, {LHCb}} 
\def\papercopyright{\the\year\ CERN for the benefit of the LHCb collaboration} 
\def\paperlicence{CC BY 4.0 licence}
\def\paperlicenceurl{https://creativecommons.org/licenses/by/4.0/}

\usepackage[top=1in, bottom=1.25in, left=1in, right=1in]{geometry}

\usepackage{microtype}
\usepackage{lineno}  
\usepackage{xspace} 
\usepackage{caption} 

\usepackage{graphicx}  
\usepackage{color}
\usepackage{colortbl}
\graphicspath{{./figs/}} 

\usepackage{amsmath} 
\usepackage{amssymb}
\usepackage{amsfonts}
\usepackage{upgreek} 

\newcommand*\patchAmsMathEnvironmentForLineno[1]{%
\expandafter\let\csname old#1\expandafter\endcsname\csname #1\endcsname
\expandafter\let\csname oldend#1\expandafter\endcsname\csname
end#1\endcsname
 \renewenvironment{#1}%
   {\linenomath\csname old#1\endcsname}%
   {\csname oldend#1\endcsname\endlinenomath}%
}
\newcommand*\patchBothAmsMathEnvironmentsForLineno[1]{%
  \patchAmsMathEnvironmentForLineno{#1}%
  \patchAmsMathEnvironmentForLineno{#1*}%
}
\AtBeginDocument{%
\patchBothAmsMathEnvironmentsForLineno{equation}%
\patchBothAmsMathEnvironmentsForLineno{align}%
\patchBothAmsMathEnvironmentsForLineno{flalign}%
\patchBothAmsMathEnvironmentsForLineno{alignat}%
\patchBothAmsMathEnvironmentsForLineno{gather}%
\patchBothAmsMathEnvironmentsForLineno{multline}%
\patchBothAmsMathEnvironmentsForLineno{eqnarray}%
}

\usepackage{hyperxmp}

\usepackage[pdftex,
            pdfauthor={\paperauthors},
            pdftitle={\paperasciititle},
            pdfkeywords={\paperkeywords},
            pdfcopyright={Copyright (C) \papercopyright},
            pdflicenseurl={\paperlicenceurl}]{hyperref}

\usepackage[colorinlistoftodos,textsize=scriptsize]{todonotes}

\usepackage[bottom,flushmargin,hang,multiple]{footmisc}

\usepackage[all]{hypcap} 

\usepackage{xspace} 
\usepackage{upgreek}

\def\lhcb   {\mbox{LHCb}\xspace}

\def\MagUp {\mbox{\em Mag\kern -0.05em Up}\xspace}

\ifthenelse{\boolean{uprightparticles}}%
{

 \def\Ppi         {\ensuremath{\uppi}\xspace}

 \def\PDelta      {\ensuremath{\Delta}\xspace}                 
 \def\PXi         {\ensuremath{\Xi}\xspace}                 
 \def\PLambda     {\ensuremath{\Lambda}\xspace}                 
 \def\PSigma      {\ensuremath{\Sigma}\xspace}                 
 \def\POmega      {\ensuremath{\Omega}\xspace}                 
 \def\PUpsilon    {\ensuremath{\Upsilon}\xspace}
 \let\oldPi\Pi
 \def\PPi         {\ensuremath{\oldPi}\xspace}

 \def\PB      {\ensuremath{\mathrm{B}}\xspace}                 
                  
 \def\PD      {\ensuremath{\mathrm{D}}\xspace}

 \def\PK      {\ensuremath{\mathrm{K}}\xspace}

 \def\Pb      {\ensuremath{\mathrm{b}}\xspace}                 
 \def\Pc      {\ensuremath{\mathrm{c}}\xspace}

 \def\Pi      {\ensuremath{\mathrm{i}}\xspace}

 \def\Ps      {\ensuremath{\mathrm{s}}\xspace}

 \def\thebaroffset{0.0em}
}
{

 \def\Ppi         {\ensuremath{\pi}\xspace}

 \mathchardef\PDelta="7101
 \mathchardef\PXi="7104
 \mathchardef\PLambda="7103
 \mathchardef\PSigma="7106
 \mathchardef\POmega="710A
 \mathchardef\PUpsilon="7107
 \mathchardef\PPi="7105
                  
 \def\PB      {\ensuremath{B}\xspace}                 
                  
 \def\PD      {\ensuremath{D}\xspace}

 \def\PK      {\ensuremath{K}\xspace}

 \def\Pb      {\ensuremath{b}\xspace}                 
 \def\Pc      {\ensuremath{c}\xspace}

 \def\Pi      {\ensuremath{i}\xspace}

 \def\Ps      {\ensuremath{s}\xspace}

 \def\thebaroffset{0.18em}
}
\newcommand{\offsetoverline}[2][\thebaroffset]{\kern #1\overline{\kern -#1 #2}}%

\makeatletter
\ifcase \@ptsize \relax
  \newcommand{\miniscule}{\@setfontsize\miniscule{4}{5}}
\or
  \newcommand{\miniscule}{\@setfontsize\miniscule{5}{6}}
\or
  \newcommand{\miniscule}{\@setfontsize\miniscule{5}{6}}
\fi
\makeatother

\DeclareRobustCommand{\optbar}[1]{\shortstack{{\miniscule (\rule[.5ex]{1.25em}{.18mm})}
  \\ [-.7ex] $#1$}}

\def\squark    {{\ensuremath{\Ps}}\xspace}

\def\cquark    {{\ensuremath{\Pc}}\xspace}

\def\bquark    {{\ensuremath{\Pb}}\xspace}

\def\pion   {{\ensuremath{\Ppi}}\xspace}

\def\pip    {{\ensuremath{\pion^+}}\xspace}
\def\pim    {{\ensuremath{\pion^-}}\xspace}

\def\kaon    {{\ensuremath{\PK}}\xspace}

\def\KorKbar {\kern \thebaroffset\optbar{\kern -\thebaroffset \PK}{}\xspace}

\def\Kp      {{\ensuremath{\kaon^+}}\xspace}
\def\Km      {{\ensuremath{\kaon^-}}\xspace}

\def\D       {{\ensuremath{\PD}}\xspace}

\def\DorDbar {\kern \thebaroffset\optbar{\kern -\thebaroffset \PD}\xspace}
\def\Dz      {{\ensuremath{\D^0}}\xspace}

\def\Dp      {{\ensuremath{\D^+}}\xspace}
\def\Dm      {{\ensuremath{\D^-}}\xspace}

\def\DpDm    {\ensuremath{\Dp {\kern -0.16em \Dm}}\xspace}

\def\Dsp     {{\ensuremath{\D^+_\squark}}\xspace}
\def\Dsm     {{\ensuremath{\D^-_\squark}}\xspace}

\def\Dsmp    {{\ensuremath{\D^{\mp}_\squark}}\xspace}

\def\B       {{\ensuremath{\PB}}\xspace}

\def\BorBbar {\kern \thebaroffset\optbar{\kern -\thebaroffset \PB}\xspace}

\def\Bd      {{\ensuremath{\B^0}}\xspace}

\def\BdorBdbar {\kern \thebaroffset\optbar{\kern -\thebaroffset \Bd}\xspace}
\def\Bu      {{\ensuremath{\B^+}}\xspace}

\def\Bp      {{\ensuremath{\Bu}}\xspace}

\def\Bs      {{\ensuremath{\B^0_\squark}}\xspace}

\def\BsorBsbar {\kern \thebaroffset\optbar{\kern -\thebaroffset \Bs}\xspace}

\def\Y#1S{\ensuremath{\PUpsilon{(#1S)}}\xspace}

\def\Lz          {{\ensuremath{\PLambda}}\xspace}

\def\LorLbar     {\kern \thebaroffset\optbar{\kern -\thebaroffset \PLambda}\xspace}

\def\Lc          {{\ensuremath{\Lz^+_\cquark}}\xspace}

\def\to                 {\ensuremath{\rightarrow}\xspace}

\def\CP                {{\ensuremath{C\!P}}\xspace}

\def\AT#1     {\ensuremath{A_{\mathrm{T}}^{#1}}\xspace}           

\def\C#1      {\ensuremath{\mathcal{C}_{#1}}\xspace}                       
\def\Cp#1     {\ensuremath{\mathcal{C}_{#1}^{'}}\xspace}                    
\def\Ceff#1   {\ensuremath{\mathcal{C}_{#1}^{\mathrm{(eff)}}}\xspace}        
\def\Cpeff#1  {\ensuremath{\mathcal{C}_{#1}^{'\mathrm{(eff)}}}\xspace}       
\def\Ope#1    {\ensuremath{\mathcal{O}_{#1}}\xspace}                       
\def\Opep#1   {\ensuremath{\mathcal{O}_{#1}^{'}}\xspace}                    

\newcommand{\nospaceunit}[1]{\ensuremath{\text{#1}}}       
\newcommand{\aunit}[1]{\ensuremath{\text{\,#1}}}       

\newcommand{\tev}{\aunit{Te\kern -0.1em V}\xspace}
\newcommand{\gev}{\aunit{Ge\kern -0.1em V}\xspace}
\newcommand{\mev}{\aunit{Me\kern -0.1em V}\xspace}
\newcommand{\kev}{\aunit{ke\kern -0.1em V}\xspace}
\newcommand{\ev}{\aunit{e\kern -0.1em V}\xspace}
\newcommand{\gevgev}{\ensuremath{\gev^2}\xspace} 
\newcommand{\mevc}{\ensuremath{\aunit{Me\kern -0.1em V\!/}c}\xspace}
\newcommand{\gevc}{\ensuremath{\aunit{Ge\kern -0.1em V\!/}c}\xspace}
\newcommand{\mevcc}{\ensuremath{\aunit{Me\kern -0.1em V\!/}c^2}\xspace}
\newcommand{\gevcc}{\ensuremath{\aunit{Ge\kern -0.1em V\!/}c^2}\xspace}

\def\mum  {\ensuremath{\,\upmu\nospaceunit{m}}\xspace}

\def\fb   {\ensuremath{\aunit{fb}}\xspace}
\def\invfb   {\ensuremath{\fb^{-1}}\xspace}

\newcommand{\chisq}{\ensuremath{\chi^2}\xspace}

\def\gsim{{~\raise.15em\hbox{$>$}\kern-.85em
          \lower.35em\hbox{$\sim$}~}\xspace}
\def\lsim{{~\raise.15em\hbox{$<$}\kern-.85em
          \lower.35em\hbox{$\sim$}~}\xspace}

\def\pt         {\ensuremath{p_{\mathrm{T}}}\xspace}

\def\ptot       {\ensuremath{p}\xspace}

\def\evtgen     {\mbox{\textsc{EvtGen}}\xspace}

\def\geant      {\mbox{\textsc{Geant4}}\xspace}

\def\photos     {\mbox{\textsc{Photos}}\xspace}

\def\pythia     {\mbox{\textsc{Pythia}}\xspace}

\def\tell1  {TELL1\xspace}
\def\ukl1   {UKL1\xspace}

\newcommand{\ie}{\mbox{\itshape i.e.}\xspace}

\newcommand{\lhcborcid}[1]{\href{https://orcid.org/#1}{\hspace*{0.1em}\raisebox{-0.45ex}{\includegraphics[width=1em]{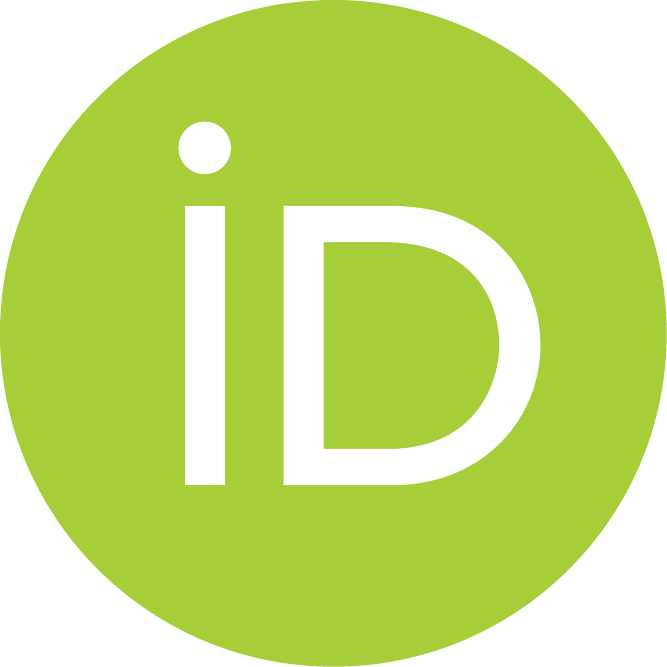}}}}

\usepackage{cite} 
\usepackage{mciteplus}

\def\Dspp     {\ensuremath{\D^+_{(\squark)}}\xspace}
\def\Dsmm     {\ensuremath{\D^-_{(\squark)}}\xspace}

\def\Dsmp

\def\DptoKKP {\ensuremath{\Dp\rightarrow \Km\Kp\pip}\xspace}

\def\DptoKPP {\ensuremath{\Dp\rightarrow \Km\pip\pip}\xspace}

\def\DptoKKK {\ensuremath{\Dp\rightarrow \Km\Kp\Kp}\xspace}
\def\DsptoKKK {\ensuremath{\Dsp\rightarrow \Km\Kp\Kp}\xspace}

\def\DsptoKKP {\ensuremath{\Dsp\rightarrow \Km\Kp\pip}\xspace}

\def\DptoKKP {\ensuremath{\Dp\rightarrow \Km\Kp\pip}\xspace}

\usepackage{rotating}
\usepackage{graphicx}
\usepackage{wrapfig}
\usepackage{multirow}
\usepackage{enumerate}
\usepackage[section]{placeins}

\usepackage{caption}
\usepackage{subcaption}
\usepackage{comment}
\begin{document}

\renewcommand{\thefootnote}{\fnsymbol{footnote}}
\setcounter{footnote}{1}


\begin{titlepage}
\pagenumbering{roman}

\vspace*{-1.5cm}
\centerline{\large EUROPEAN ORGANIZATION FOR NUCLEAR RESEARCH (CERN)}
\vspace*{1.5cm}
\noindent
\begin{tabular*}{\linewidth}{lc@{\extracolsep{\fill}}r@{\extracolsep{0pt}}}
\ifthenelse{\boolean{pdflatex}}
{\vspace*{-1.5cm}\mbox{\!\!\!\includegraphics[width=.14\textwidth]{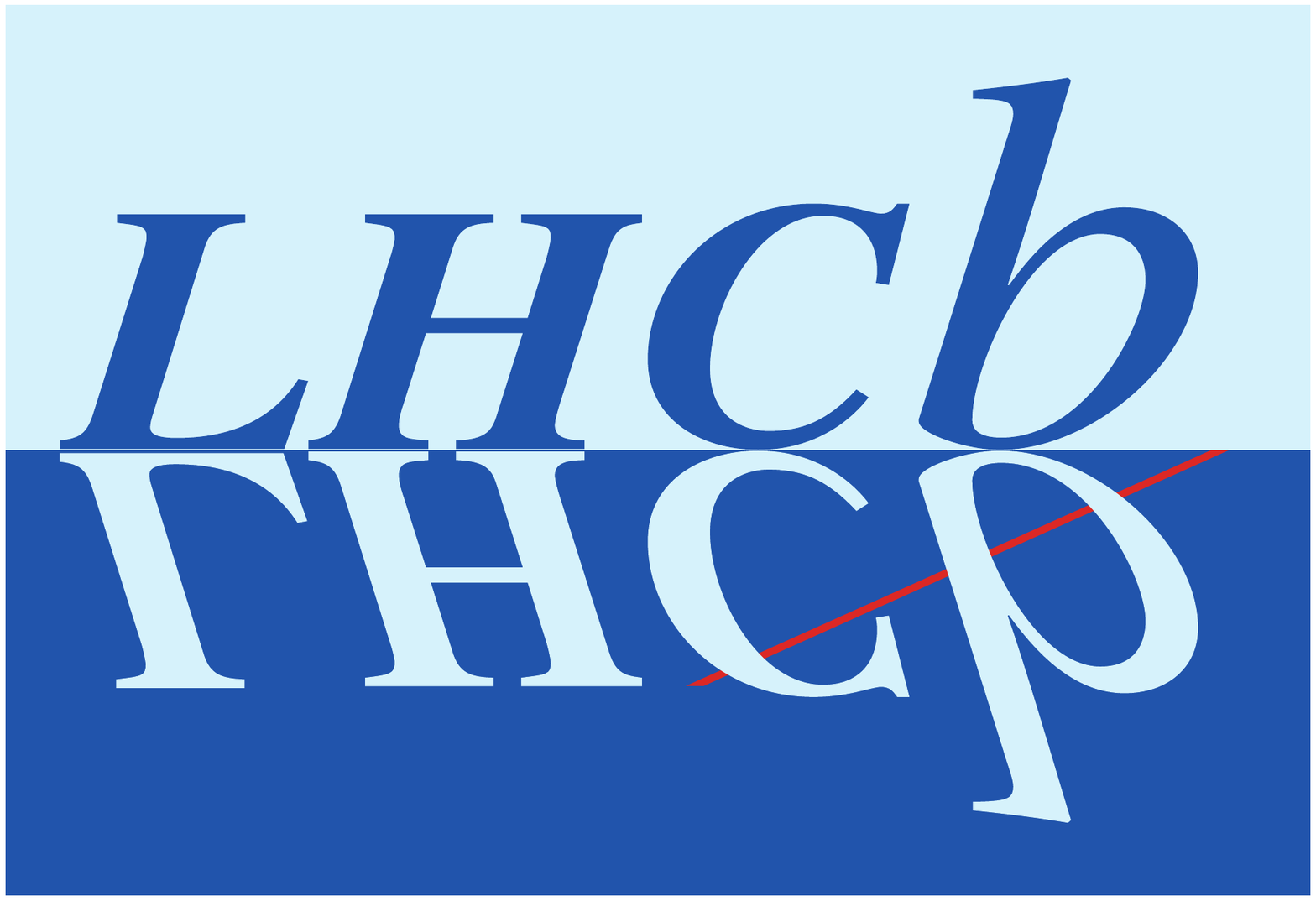}} & &}%
{\vspace*{-1.2cm}\mbox{\!\!\!\includegraphics[width=.12\textwidth]{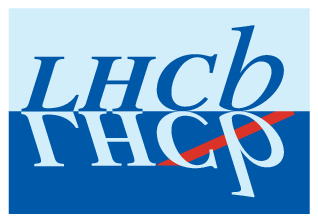}} & &}%
\\
 & & CERN-EP-2023-010 \\  
 & & LHCb-PAPER-2022-042 \\  
 & & 17 July 2023 \\ 
 & & \\
\end{tabular*}

\vspace*{3.5cm}

{\normalfont\bfseries\boldmath\huge
\begin{center}
  \papertitle 
\end{center}
}

\vspace*{2.0cm}

\begin{center}
\paperauthors\footnote{Authors are listed at the end of this paper.}
\end{center}

\vspace{\fill}

\begin{abstract}
  \noindent
  A search for direct \CP violation in the Cabibbo-suppressed decay \DsptoKKK and in  the doubly Cabibbo-suppressed decay \DptoKKK  is reported. The analysis is performed with data collected by the LHCb experiment  in proton-proton collisions at a centre-of-mass energy of 13\tev corresponding to an integrated luminosity of 5.6\invfb. The search is conducted by comparing the $D^+_{(s)}$ and $D^-_{(s)}$ Dalitz-plot distributions through a model-independent binned technique, based on fits to the $\Km\Kp\Kp$ invariant-mass distributions, with a total of 0.97 (1.27) million \Dsp (\Dp)  signal candidates. 
  The results are given as $p$-values for the hypothesis of \CP conservation and are found to be 13.3\% for the \DsptoKKK decay and 31.6\% for the \DptoKKK decay. No evidence for  \CP violation is observed in these decays.
  
\end{abstract}

\vspace*{2.0cm}

\begin{center}
  Published in JHEP 07 (2023) 067
\end{center}

\vspace{\fill}

{\footnotesize 
\centerline{\copyright~\papercopyright. \href{\paperlicenceurl}{\paperlicence}.}}
\vspace*{2mm}

\end{titlepage}


\newpage
\setcounter{page}{2}
\mbox{~}
%
%
%
%


\renewcommand{\thefootnote}{\arabic{footnote}}
\setcounter{footnote}{0}

\cleardoublepage


\pagestyle{plain} 
\setcounter{page}{1}
\pagenumbering{arabic}



\section{Introduction}
\label{sec:intro}

Since its first observation almost six decades ago, charge-parity (\CP) violation has been extensively studied, with clear manifestation in decays of $K$ and $B$ mesons. In the Standard Model~(SM), \CP violation (CPV) is realised through the complex phase of the Cabibbo-Kobayashi-Maskawa~(CKM) matrix \cite{Kobayashi:1973fv}. For  the charm sector, given the CKM hierarchy~\cite{PDG2022}, CPV effects are  very small: direct CPV (\ie in the decay) can  occur only for
Cabibbo-suppressed~(CS) decays for which  \CP asymmetries  are expected at  $\mathcal{O}(10^{-3})$ or less \cite{Grossman_CPV_CS_2007,Bianco_Bigi_2020,Grossman_2019,Lenz_CHARM2013,Khodjamirian_Petrov_2017,Golden_Grinstein_1989,Cheng_Chiang_2012_1,Brod_ConsistentPenguins_2012,Zhang:2021zhr,Schacht:2021jaz,Bediaga:2022sxw,Li_ImplicationsCPV_2019,Lenz:2020awd,Buccella_2019,Cheng_2019,Cheng:2021yrn}. For doubly Cabibbo-suppressed~(DCS) decays, CPV is essentially forbidden, thus its observation would indicate a manifestation of physics beyond the SM (BSM)~\cite{Bergmann:1999pm,Grossman_CPV_CS_2007,Bianco_Bigi_2020}.

To date there is only one observation of CPV in the charm sector, through the difference of \CP asymmetries in $\Dz\to \Km\Kp$ and $\Dz\to \pim\pip$ decays~\cite{LHCb-PAPER-2019-006}, which is mainly a direct CPV observable~\cite{Gersabeck_Interplay_2012,HFLAV2022}. New evidence indicates the main source for this CPV effect to be from the $\Dz\to \pim\pip$ channel~\cite{LHCb-PAPER-2022-024}. 
Although the measured \CP asymmetry,   at the $0.1\%$ level,  is consistent  with some estimates within the SM \cite{Li_ImplicationsCPV_2019,Zhang:2021zhr,Schacht:2021jaz,Bediaga:2022sxw} 
other estimates indicate that BSM contributions are required to explain this result
\cite{Nir_Implications_CharmCPV_2019,Chala_2019,Calibbi_2019}. It is thus important to extend the studies to different charm-hadron species decaying into a broad range of final states, including not only CS but also DCS decays~\cite{Bianco_Bigi_2020}.

Direct CPV occurs when a given final state is produced through amplitudes with  different weak phases while the presence of different strong phases is also required~\cite{Branco_Book,*Bigi_Book}. Three-body charm decays are particularly interesting in this context: final states are reached 
mainly through resonances, providing an important source of strong-phase differences that vary across the two-dimensional Dalitz plot~\cite{Dalitz:1953cp,*Fabri:1954zz}. 
This feature can enhance the sensitivity to \CP asymmetries in localised regions of the phase space when compared to global charge asymmetries (\ie integrated over the Dalitz plot)~\cite{bigi3,Bigi_CP-2-3-4Body_2011,Bigi_Charm_2015,BediagaGobel} --- as   observed in charmless \Bp decays~\cite{LHCb-PAPER-2021-049}. 
Compared to $B$ decays, however, the expected effects for charm are very small and there is no indication on how they would appear throughout the Dalitz plot. For this reason, recent studies usually make use of model-independent methods either in bins of the Dalitz plot ~\cite{mirandizing2,CDF-D0KSpipi-2012,LHCb-PAPER-2011-017,LHCb-PAPER-2013-057,LHCb-PAPER-2013-041,LHCb-PAPER-2019-026} or via unbinned techniques~\cite{LHCb-PAPER-2013-041, LHCb-PAPER-2014-054, LHCb-PAPER-2016-044, LHCb-PAPER-2019-026} to search for signs of \CP violation and establish a \textit{p}-value for the hypothesis of \CP conservation.

This paper describes a model-independent search for direct CPV in the Dalitz plots of the CS \DsptoKKK and DCS \DptoKKK decays\footnote{The inclusion of charge-conjugate processes is implied throughout, unless otherwise stated.}
with data collected by the LHCb experiment from 2016 to 2018 in proton-proton ($pp$) collisions at a centre-of-mass energy $\sqrt{s} = 13 \tev$ corresponding to an integrated luminosity of 5.6\invfb.
The applied search method is a variation of the original {\it Miranda} technique \cite{mirandizing2,mirandizing1}, which has been used in several studies~\cite{LHCb-PAPER-2011-017,LHCb-PAPER-2013-041,LHCb-PAPER-2013-057,LHCb-PAPER-2014-020,LHCb-PAPER-2019-026}. The method involves dividing the Dalitz plot  in two-dimensional bins and computing, for each bin,  the significance of the difference in the numbers of \Dspp candidates and \Dsmm candidates, where the latter is corrected for global charge asymmetry. A two-sample $\chi^2$ test is then performed on the \Dspp and \Dsmm samples. In the original Miranda approach, these numbers are obtained by counting the candidates within a defined $\Km\Kp\Kp$  invariant-mass signal region, including background. In the approach used here, instead, the yields of \Dspp and \Dsmm\ signal candidates are obtained in each Dalitz-plot bin by a fit to the $\Km\Kp\Kp$ invariant-mass spectrum of candidates lying in that bin, thus background candidates are not counted.  
Given the number of degrees of freedom, the probability ($p$-value) of observing a $\chi^2$ greater than the measured value is computed under the null hypothesis, which in this case is \CP conservation.  
The method is insensitive to global sources of charge asymmetries, but relies on  the absence of residual local nuisance asymmetries, for instance from production and detection.
This assumption is validated using simulation, by studying the background in the $\Km\Kp\Kp$ samples, and by analysing large samples of the Cabibbo-favoured modes \DsptoKKP and \DptoKPP for which CPV is not allowed.

\section{LHCb detector}
\label{sec:detector}

The \lhcb detector~\cite{LHCb-DP-2008-001,LHCb-DP-2014-002} is a single-arm forward
spectrometer covering the \mbox{pseudorapidity} range $2<\eta <5$,
designed for the study of particles containing \bquark or \cquark
quarks. The detector includes a high-precision tracking system
consisting of a silicon-strip vertex detector surrounding the $pp$
interaction region, a large-area silicon-strip detector located
upstream of a dipole magnet with a bending power of about
$4{\mathrm{\,Tm}}$, and three stations of silicon-strip detectors and straw
drift tubes placed downstream of the magnet. The magnetic field deflects oppositely charged particles in opposite
directions which can lead to detection asymmetries. Periodically
reversing the magnetic field polarity throughout  data-taking cancels most of this  effect. 
The tracking system provides a measurement of the momentum, \ptot, of charged particles with
a relative uncertainty that varies from 0.5\% at low momentum to 1.0\% at 200\gev.\footnote{Natural units
      with $\hbar=c=1$ are used throughout.}
The minimum distance of a track to a primary $pp$ collision vertex (PV), the impact parameter (IP), 
is measured with a resolution of $(15+29/\pt)\mum$,
where \pt is the component of the momentum transverse to the beam, in\,\gev.
Different types of charged hadrons are distinguished using information
from two ring-imaging Cherenkov detectors. 
Photons, electrons and hadrons are identified by a calorimeter system consisting of
scintillating-pad and preshower detectors, and an electromagnetic
and a hadronic calorimeter. Muons are identified by a
system composed of alternating layers of iron and multiwire
proportional chambers.
The online event selection is performed by a trigger, which consists of a hardware stage followed by a two-level software stage. In between the two software stages, an alignment and calibration of the detector is performed in near real-time and their results are used in the trigger~\cite{LHCb-PROC-2015-011}.
      This offers the opportunity to perform physics analyses 
      directly using candidates reconstructed in the trigger 
      \cite{LHCb-DP-2012-004,LHCb-DP-2016-001} 
      which the present analysis exploits. Only the reconstructed quantities of triggered candidates are stored, which reduces the event size by an order of magnitude. 
Trigger decisions at the hardware stage are associated with reconstructed particles, and the decisions with associations are available in the offline selection. Hence requirements are made on the trigger decision on whether it was due to the signal candidates, due to other particles in the $pp$ collision, or a combination of both.

 Simulation is used to model the shape of the signal  $\Km\Kp\Kp$ mass distribution accounting for the effects of the detector acceptance, reconstruction and selection criteria. 
  In the simulation, $pp$ collisions are generated using
  \pythia~\cite{Sjostrand:2007gs,Sjostrand:2006za} 
  with a specific \lhcb configuration~\cite{LHCb-PROC-2010-056}.
  Decays of unstable particles
  are described by \evtgen~\cite{Lange:2001uf}, in which final-state
  radiation is generated using \photos~\cite{davidson2015photos}.
  The interaction of the generated particles with the detector, and its response,
  are implemented using the \geant
  toolkit~\cite{Allison:2006ve, *Agostinelli:2002hh} as described in
  Ref.~\cite{LHCb-PROC-2011-006}. 
  The underlying $pp$ interaction is reused multiple times, with an independently generated signal decay for each~\cite{LHCb-DP-2018-004}.
  
  \section{Candidate selection and final data samples}
  \label{sec:data}
  
  The \DptoKKK and \DsptoKKK decay candidates are selected online by a dedicated software trigger. Three charged particles identified as kaons are combined to form  a good-quality decay vertex, detached from any PV. The PV with the smallest value of $\chi^2_{\rm IP}$ is associated to the decay candidate, where $\chi^2_{\rm IP}$ is defined as the difference in the vertex-fit \chisq of the PV reconstructed with and without the particle under consideration, in this case the \Dspp candidate. Further requirements are applied on the $D^{+}_{(s)} $ decay time; on the angle between the reconstructed $D^{+}_{(s)}$ momentum vector and the vector connecting the PV to the decay vertex; on the $\chi^2$ of the $D^{+}_{(s)}$ decay vertex fit;  and on the momentum, the transverse momentum and the $\chi^2_{\rm IP}$ of the $D^{+}_{(s)}$ candidate and its decay products. The invariant masses of the \Dp and \Dsp candidates are required to be within the interval \mbox{1805--1935}\mev and \mbox{1905--2035}\mev, respectively.
  
  In the offline selection, further  particle identification (PID) requirements are applied to the decay products  to reduce the  cross-feed from other charm-hadron decays. For the \mbox{\DsptoKKK} channel, the most significant contribution comes from the \mbox{\DptoKKP} decay, which is reduced by a stringent PID requirement on the two same-charge kaons. This  also reduces  contributions from $\Lc \to K^-K^+p$ and $\Lc \to K^-\pi^+p$ decays. For the \DptoKKK channel, the only significant cross-feeds are the aforementioned \Lc decays, which are sufficiently reduced by the PID requirements that maximise the statistical significance for the signal.  

A small fraction of \Dspp candidates is rejected when 
two tracks or segments of tracks are duplicates of each other, resulting in reconstructed same-charge kaon candidates with very similar momenta.
Fiducial requirements are applied in order to remove small regions of the phase space where large charge asymmetries are observed, typically caused by low-momentum tracks being swept out of the detector acceptance by the magnetic field.

The remaining background, mostly due to random combinations of tracks, is reduced by using a boosted decision tree (BDT) classifier~\cite{Breiman,AdaBoost}, as implemented in the TMVA
toolkit~\cite{Hocker:2007ht,TMVA4}. 
Only quantities related to the three-track combinations are used as inputs for  the BDT classifier. In addition to quantities used in the  online selection, such as flight direction, the quantities IP, $\chi^2_{\rm IP}$, the distance between the PV and the $D^{+}_{(s)}$ decay vertex and its significance are also used. 
Approximately 3\% of the data  is used to train the algorithm and discarded for the rest of the analysis. The signal and the background samples are obtained from this fraction of the data via the {\it sPlot} technique~\cite{Pivk:2004ty} using the $\Km\Kp\Kp$ invariant mass as discriminating variable. Separate BDT classifiers  are trained for the \DsptoKKK and \DptoKKK samples. The requirements on the BDT outputs are chosen to maximise the statistical significance of the signal yields.

\begin{figure}[tb]
  \begin{center}
    \includegraphics[width=0.49\linewidth]{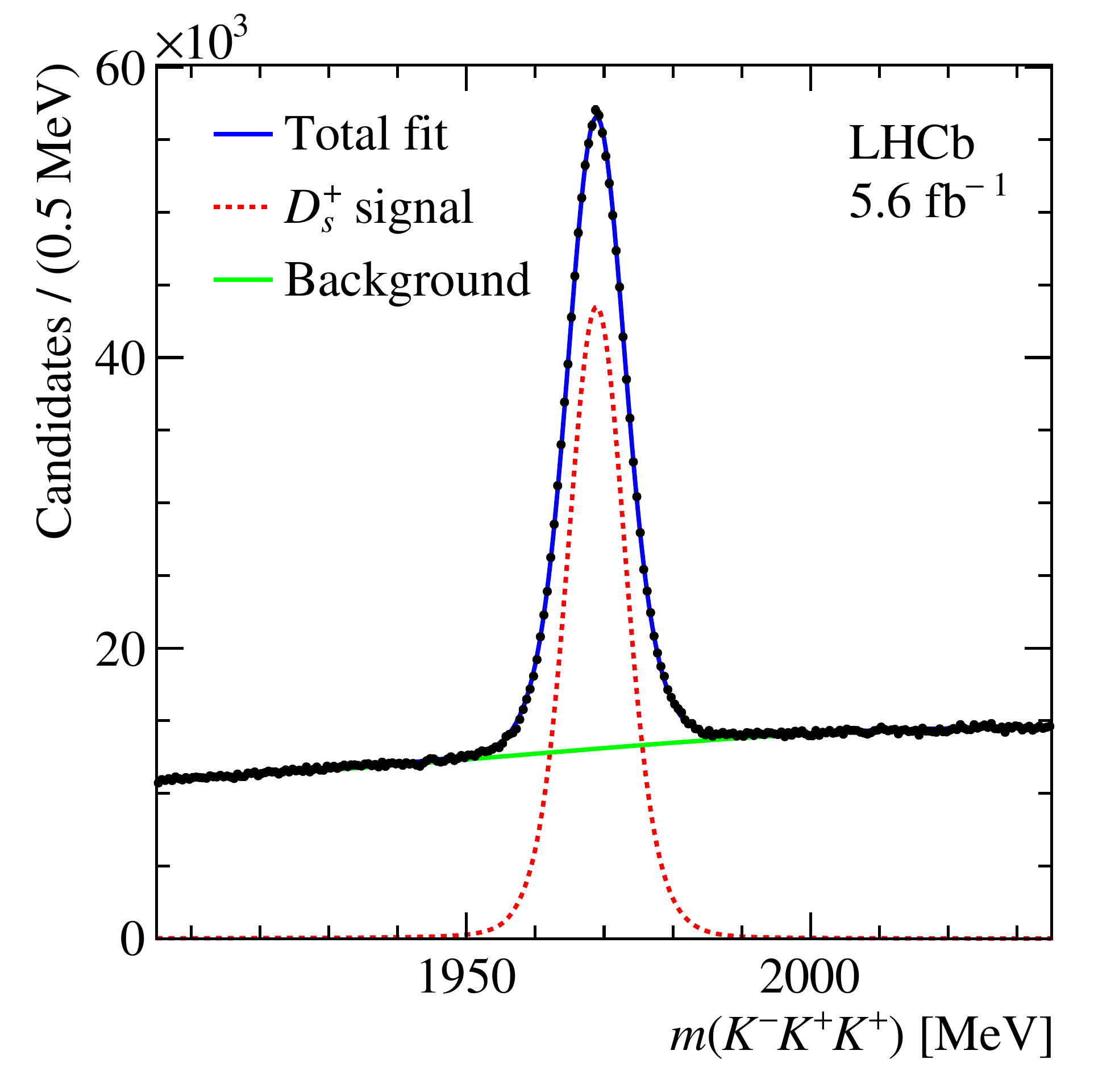}
    \includegraphics[width=0.49\linewidth]{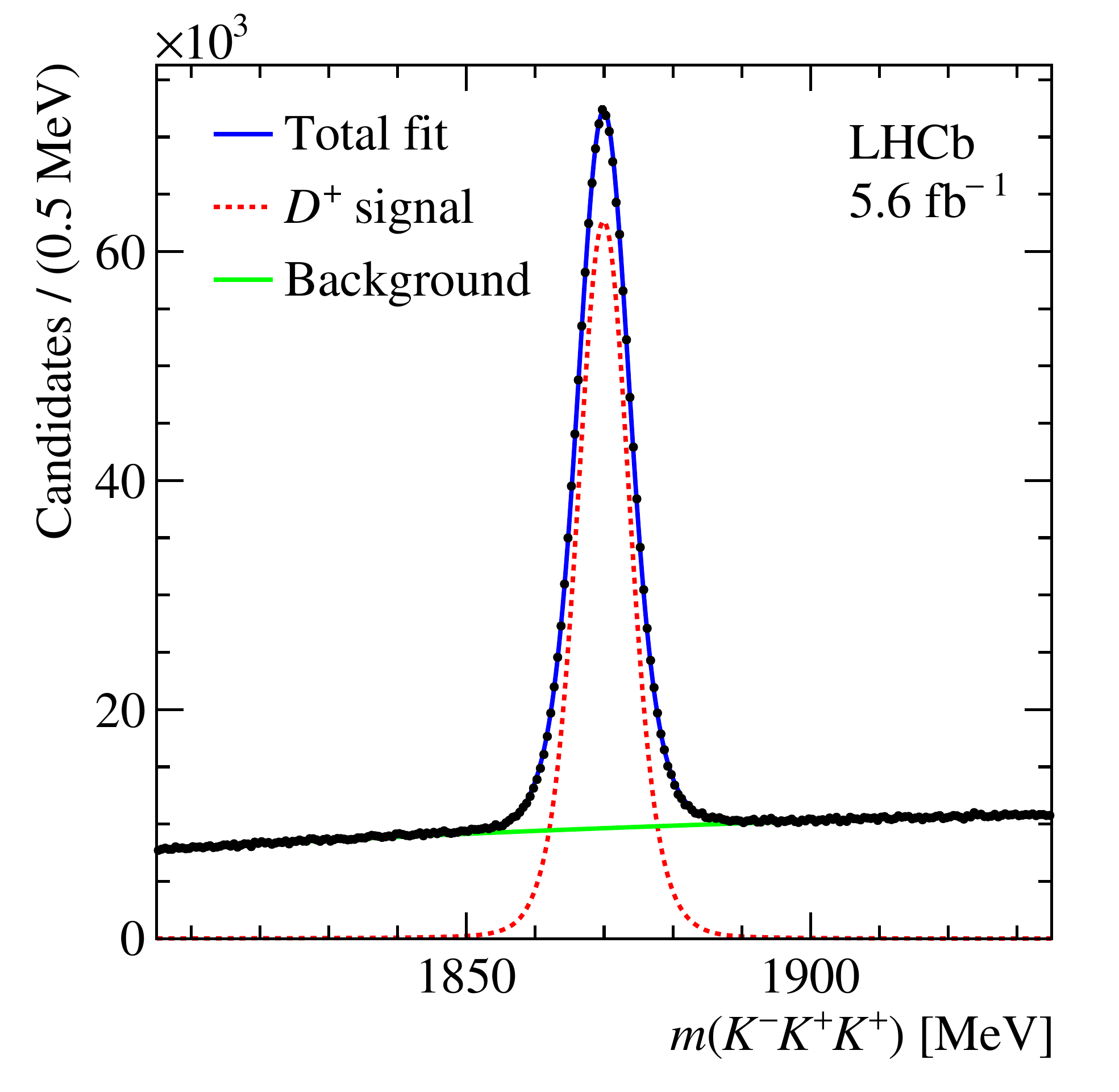}
    \vspace*{-0.5cm}
  \end{center}
  \caption{
    \small Invariant-mass distributions for (left) \DsptoKKK and (right) \DptoKKK candidates. The data are shown as points with the fit overlaid.
    }
  \label{fig:mass-fit}
\end{figure}

The $\Km\Kp\Kp$ invariant-mass distributions for the final samples are shown in Fig.~\ref{fig:mass-fit}, with the results of maximum-likelihood fits overlaid. The signal shape is described by a Gaussian and two Crystal-Ball~\cite{crystalball} functions (CB) 
with a common mean. Each CB has a width defined as a factor times the Gaussian width. The CB tails are on opposite sides.
All the parameters describing the signal shape, apart from the Gaussian width and mean, are fixed to the values obtained from simulation, including the relative fraction of each component. The background is parameterised by a third-order Bernstein polynomial.
The selection results in signal yields of 0.97 million  \DsptoKKK decays and 1.27 million \mbox{\DptoKKK} decays. Roughly 1\% of the events contain multiple signal candidates.
The purity of the samples in a region comprising 95\% of the signal candidates is about $64\%$ and 78\% for  \Dsp and \Dp decays, respectively.

The Dalitz plots are defined in terms of the variables $s_{\mathrm{high}}$ and $s_{\mathrm{low}}$, which represent the higher and lower values of the squared invariant masses formed by the two $\Km\Kp$ combinations. 
The momenta used to compute these quantities are obtained from a kinematic fit~\cite{Hulsbergen:2005pu} in which the invariant mass of the reconstructed candidates is constrained
to the known \Dspp mass~\cite{PDG2022}.  These Dalitz plots can be seen in Fig.~\ref{fig:21bingrids} for candidates within the $\Km\Kp\Kp$ invariant-mass region comprising 95\% of the signal candidates. For both decay modes the contribution of the $\phi(1020)K^+$ channel is visible. An amplitude analysis was performed recently for the \DptoKKK decay~\cite{LHCb-PAPER-2018-039} where the  $f_0(980)K^+ $ and $f_0(1370)K^+$ channels were also found to contribute. To date, no amplitude analysis exists for the \DsptoKKK decay, but its Dalitz distribution seems to follow qualitatively the same pattern of that of the \Dp decay: a clear $\phi(1020)$ signature plus a rather smooth distribution elsewhere.

\section{Method}
\label{sec:method}

The binned model-independent technique used in this analysis compares the Dalitz-plot distributions for particles and antiparticles, and it is  a variation of the original Miranda technique~\cite{mirandizing1,mirandizing2}. For each Dalitz-plot bin, 
the local \CP observable $\mathcal{S}_{\CP}$ is defined as\footnote{This expression is the same as used in previous publications~\cite{LHCb-PAPER-2013-041,LHCb-PAPER-2013-057,LHCb-PAPER-2019-026} and although it differs in the denominator from that in Ref.~\cite{mirandizing2}, for  $\alpha$ close to one the values of $\mathcal{S}_{\CP}$ are nearly identical.}
\begin{equation}
\mathcal{S}_{\CP}^i = \frac{N^i(\Dspp)-\alpha N^i(\Dsmm)}
{\sqrt{\alpha(\delta^2_{N^i(\Dspp)}+\delta^2_{N^i(\Dsmm)})}} \ , \hskip .5cm  
{\rm with} ~\alpha=\frac{\sum_i N^i(\Dspp)}{\sum_i N^i(\Dsmm)}\ ,
\label{eq:scp1}
\end{equation}
where $N^i(D^+_{(s)})$ and $N^i(D^-_{(s)})$ represent the number of 
signal candidates in the $i^{th}$ bin of the $D^+_{(s)}$ and $D^-_{(s)}$ Dalitz plots, respectively, 
and $\delta_N^i(\Dspp)$, $\delta_N^i(\Dsmm)$ their corresponding statistical uncertainties. The factor $\alpha$
normalises the total (Dalitz-plot integrated) \Dsmm yield to that of \Dspp.
The method is therefore only sensitive to local \CP asymmetries, while global effects such as production and detection asymmetries, if uniform across the Dalitz plot, are corrected for. In the absence of CPV, the values of $\mathcal{S}_{\CP}^i$ follow a standard normal distribution. A two-sample $\chi^2$ test is conducted, with $\chi^2=\sum (\mathcal{S}_{\CP}^i)^2$, and with the number of degrees of freedom (ndof) being the number of bins minus one 
(due to the constraint of the $\alpha$ normalisation).   The resulting \textit{p}-value from this test is defined as the probability of obtaining  a $\chi^{2}$ that is at least as high as the value observed, under the assumption of \CP conservation (null hypothesis). 
The criterion used is that \CP violation is  observed for \textit{p}-values less than $3\times 10^{-7}$. 

The  original Miranda method relies on counting the number of particle and antiparticle candidates in each bin, including contributions from both signal and background, with corresponding uncertainties given by their square roots. 
In this paper, a novel approach is introduced, in which the test is performed by obtaining $N^i(D^\pm_{(s)})$ and corresponding uncertainties through fits to the $K^\mp K^\pm K^\pm$ invariant-mass distribution of the candidates in each bin. By doing so, the effect of the background contribution to the calculation of $\mathcal{S}_{\CP}$ is removed. This is particularly relevant for large samples with a significant level of background: when a source of global charge asymmetry (such as that from production effects) affects the signal and the background differently, the $\alpha$ factor calculated using the original Miranda method, which includes contributions from both signal and background, may introduce a bias as shown  below.

\begin{figure}[tb]
  \begin{center}
    \includegraphics[width=0.49\linewidth]{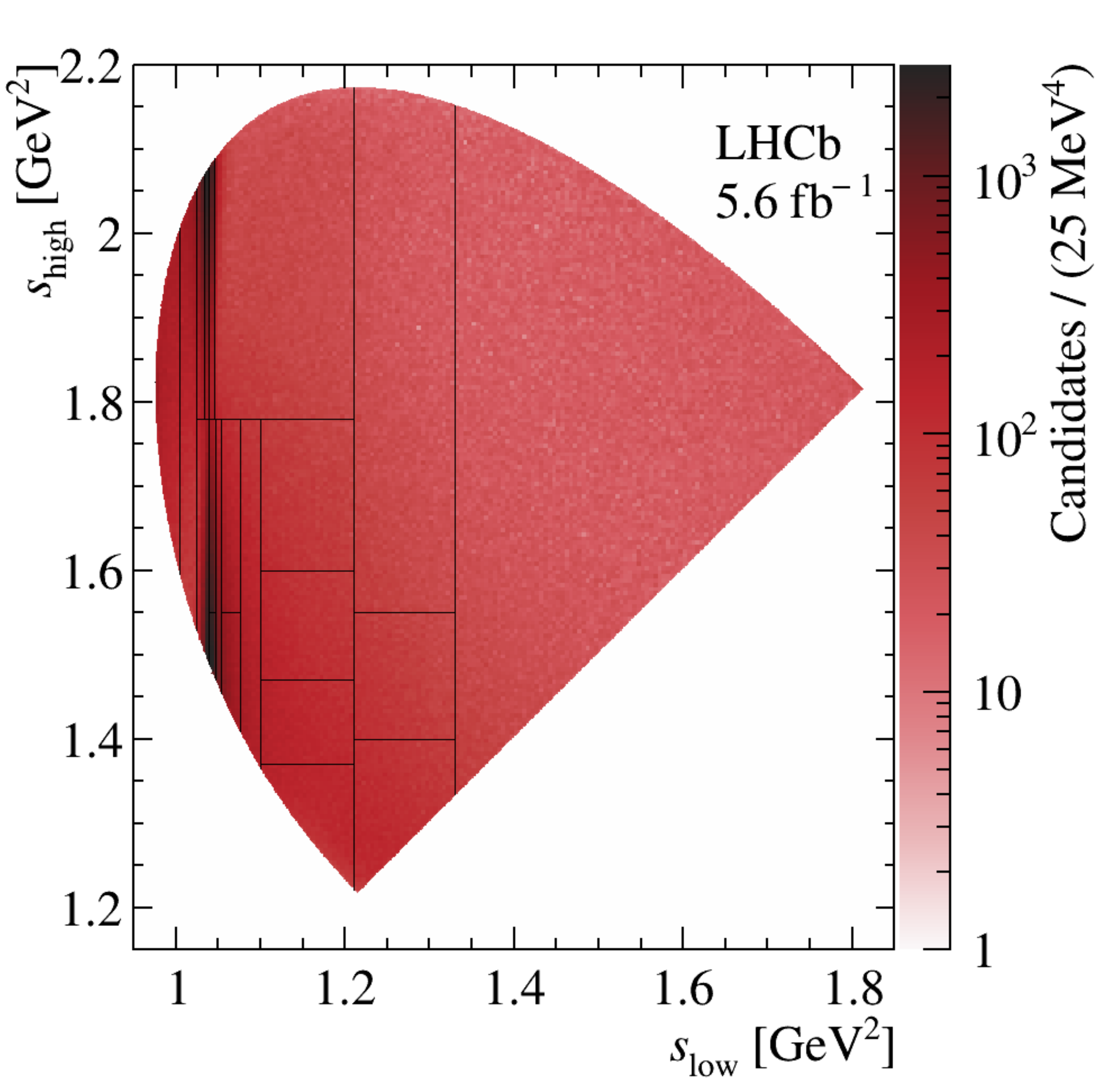}
    \includegraphics[width=0.49\linewidth]{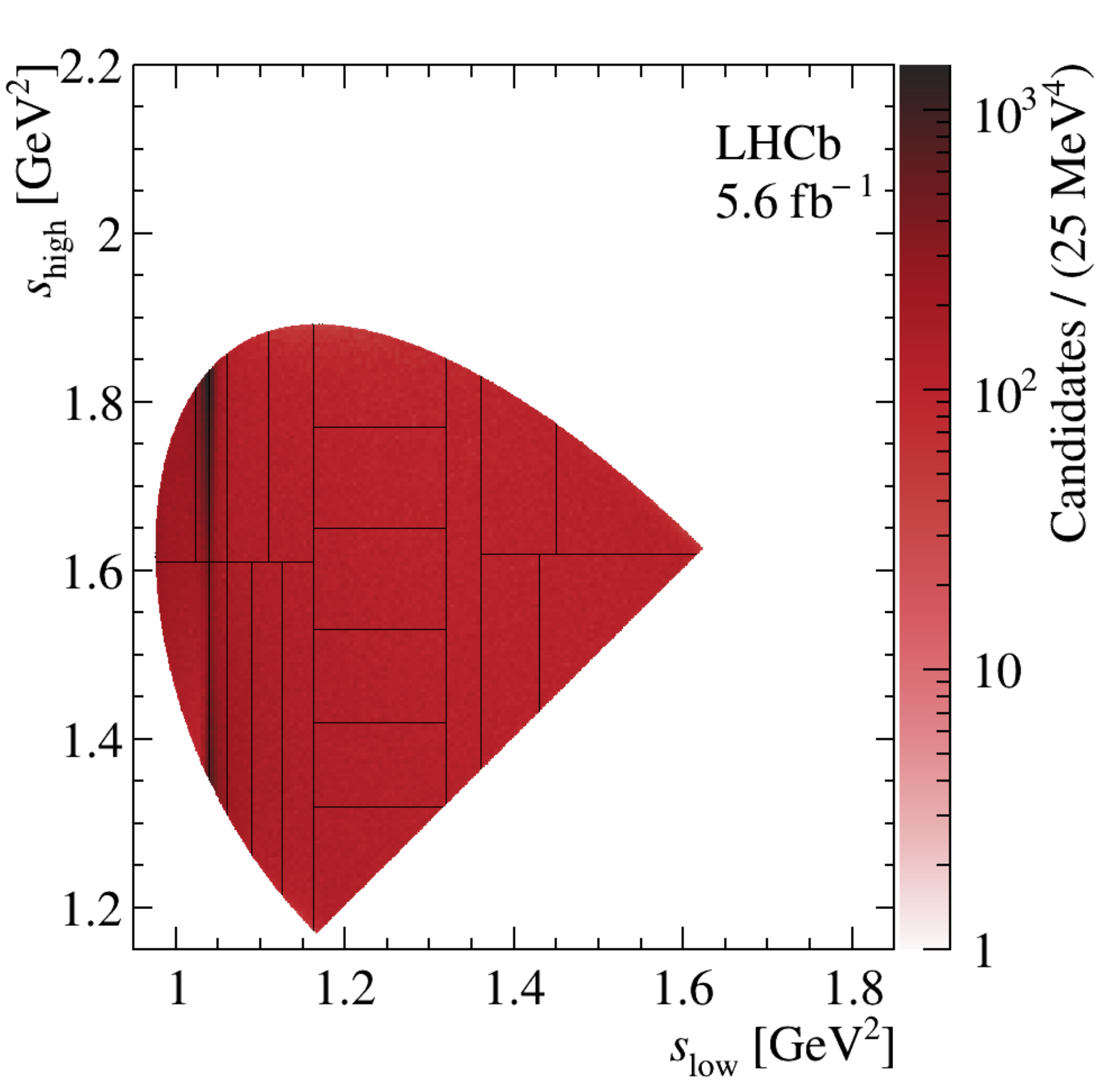}
     \caption{
    \small Dalitz plot distribution of (left) \DsptoKKK and (right) \DptoKKK  candidates, within a $\Km\Kp\Kp$ mass region comprising 95\% of the total amount of signal candidates. The binning scheme with 21 bins is overlaid in each case. The Dalitz plots are displayed within the same ranges for $s_{\rm low}$ and $s_{\rm high}$ for a better comparison.  }
  \label{fig:21bingrids}
  \end{center}
\end{figure}

\begin{figure}[tb]
  \begin{center}
    \includegraphics[width=0.45\linewidth]{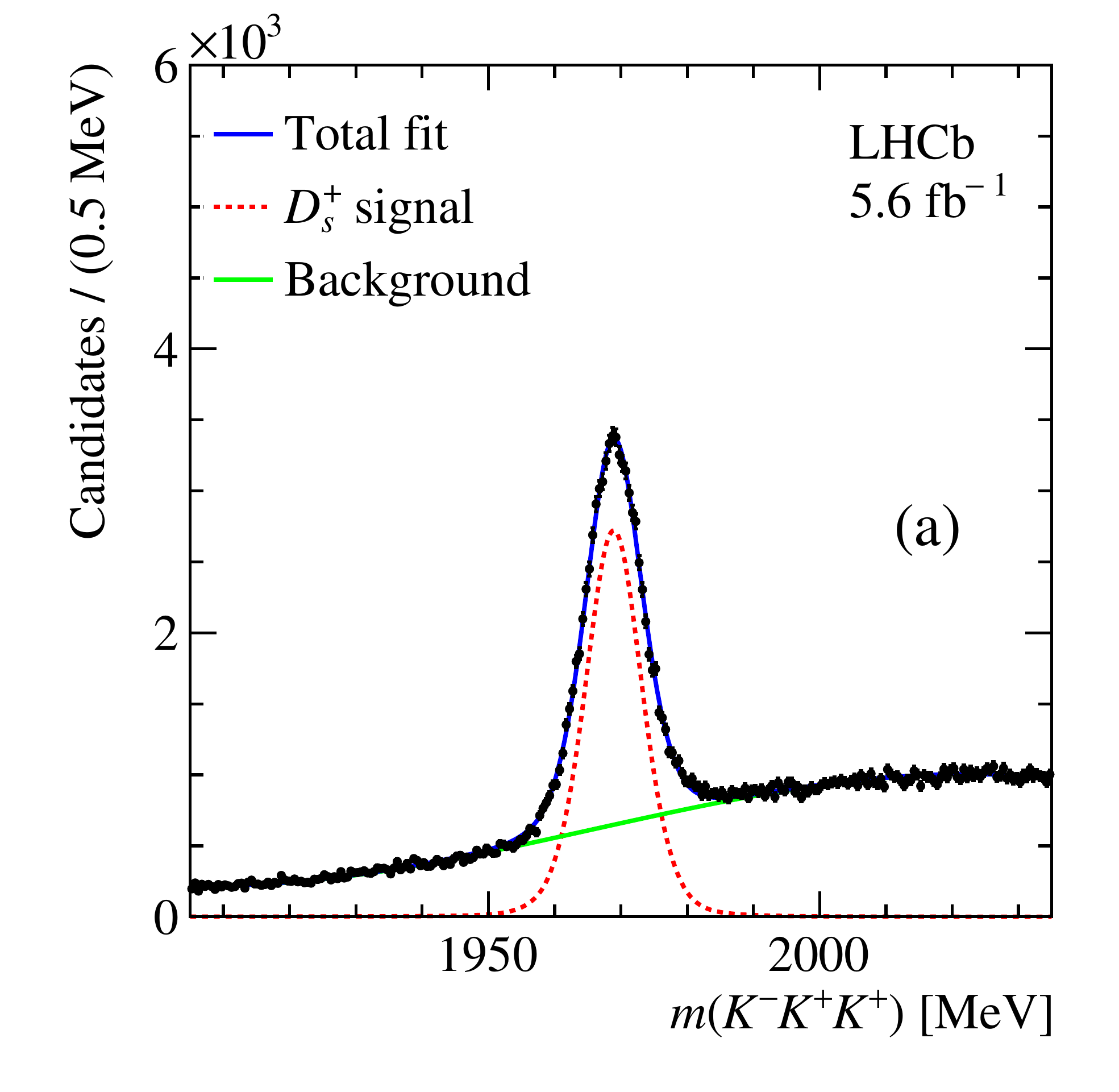}
    \includegraphics[width=0.45\linewidth]{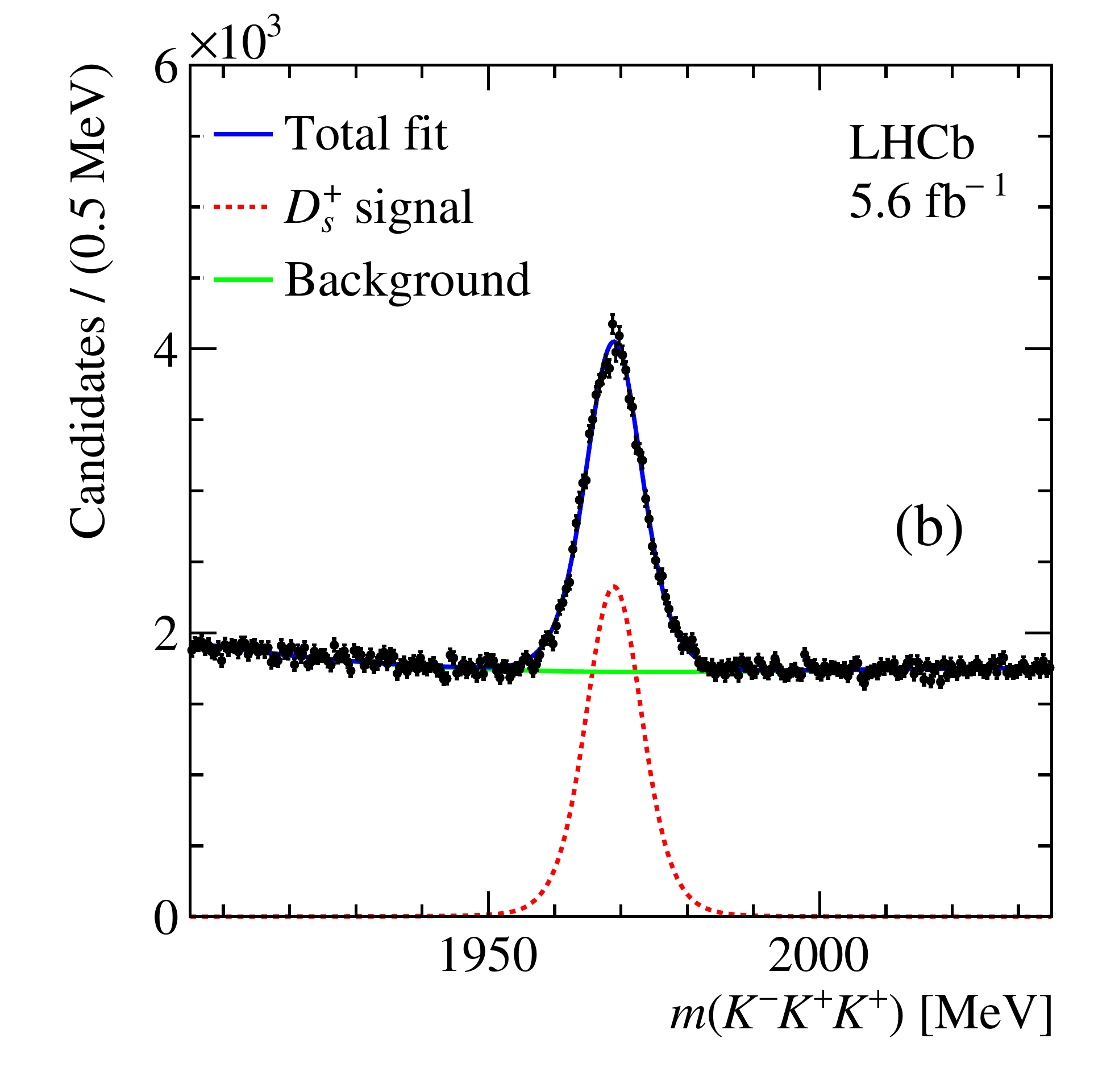}
    
    \includegraphics[width=0.45\linewidth]{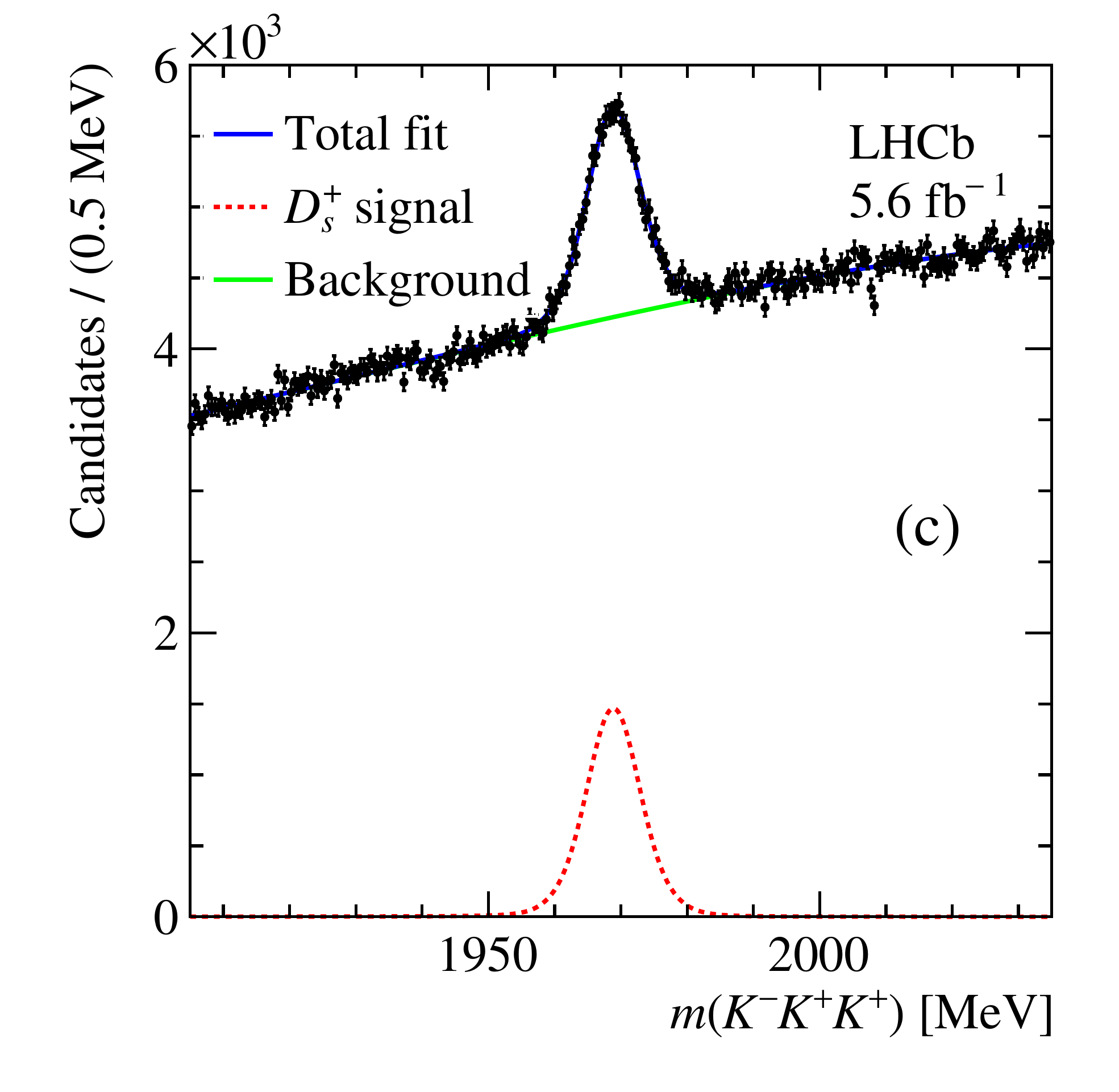}
    \includegraphics[width=0.45\linewidth]{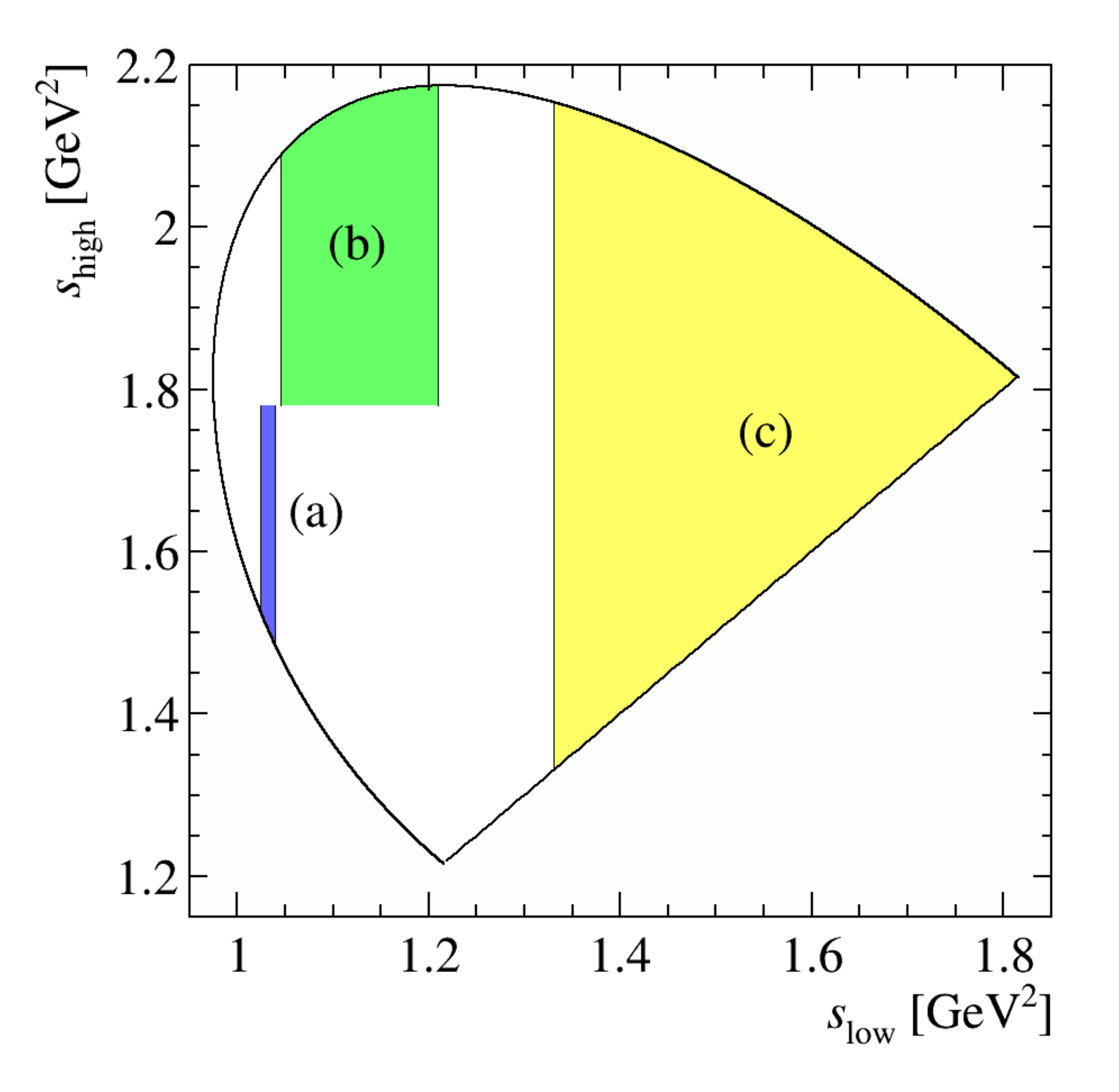}
    \vspace*{-0.5cm}
  \end{center}
  \caption{
    \small Mass distributions for the $\Dsp$ candidates in three representative Dalitz plot bins (a,b,c), defined in the lower right subfigure.}
  \label{fig:dskkk-binfit}
\end{figure}

\begin{figure}[tb]
  \begin{center}
    \includegraphics[width=0.45\linewidth]{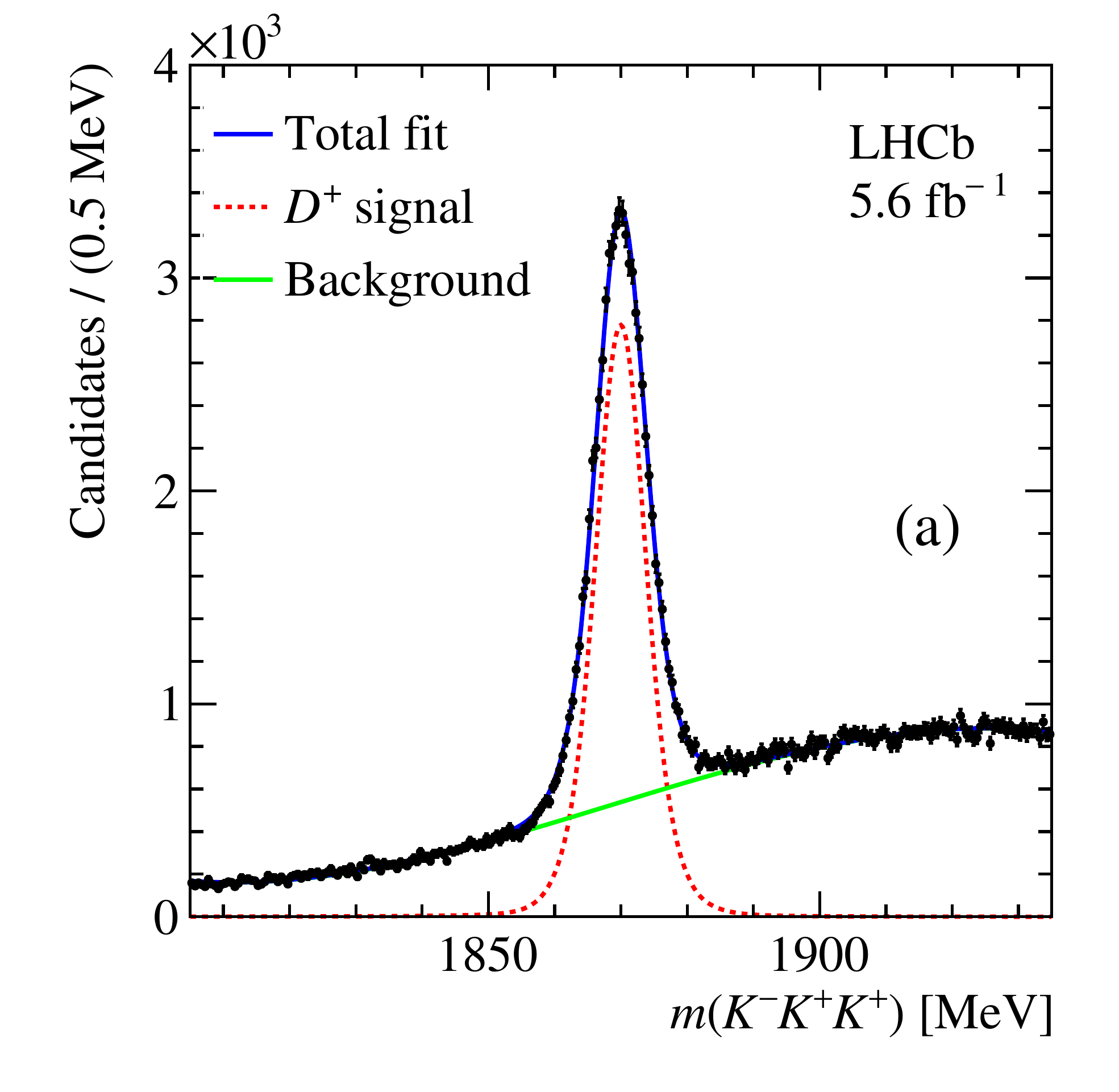}
    \includegraphics[width=0.45\linewidth]{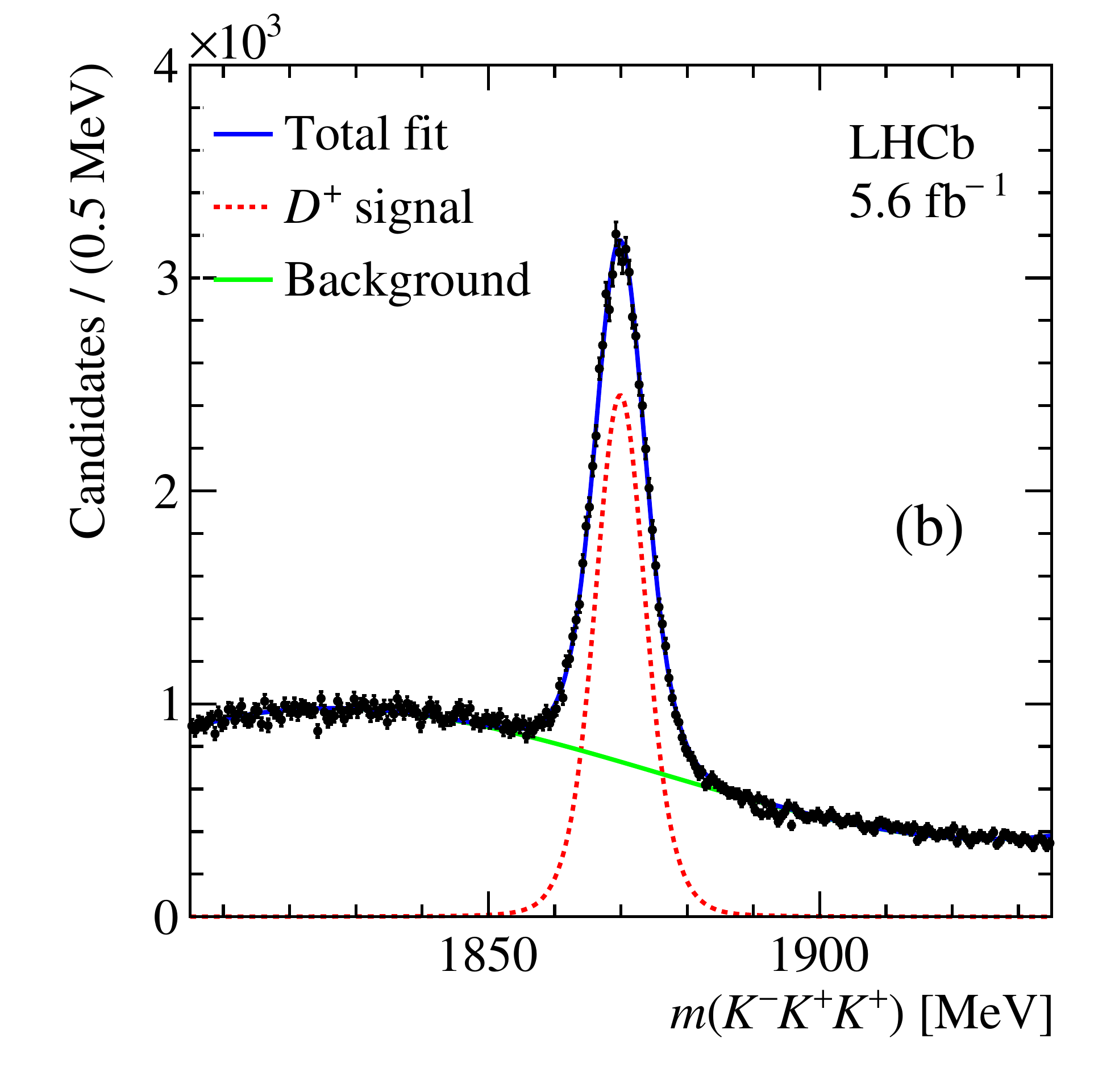}
    
    \includegraphics[width=0.45\linewidth]{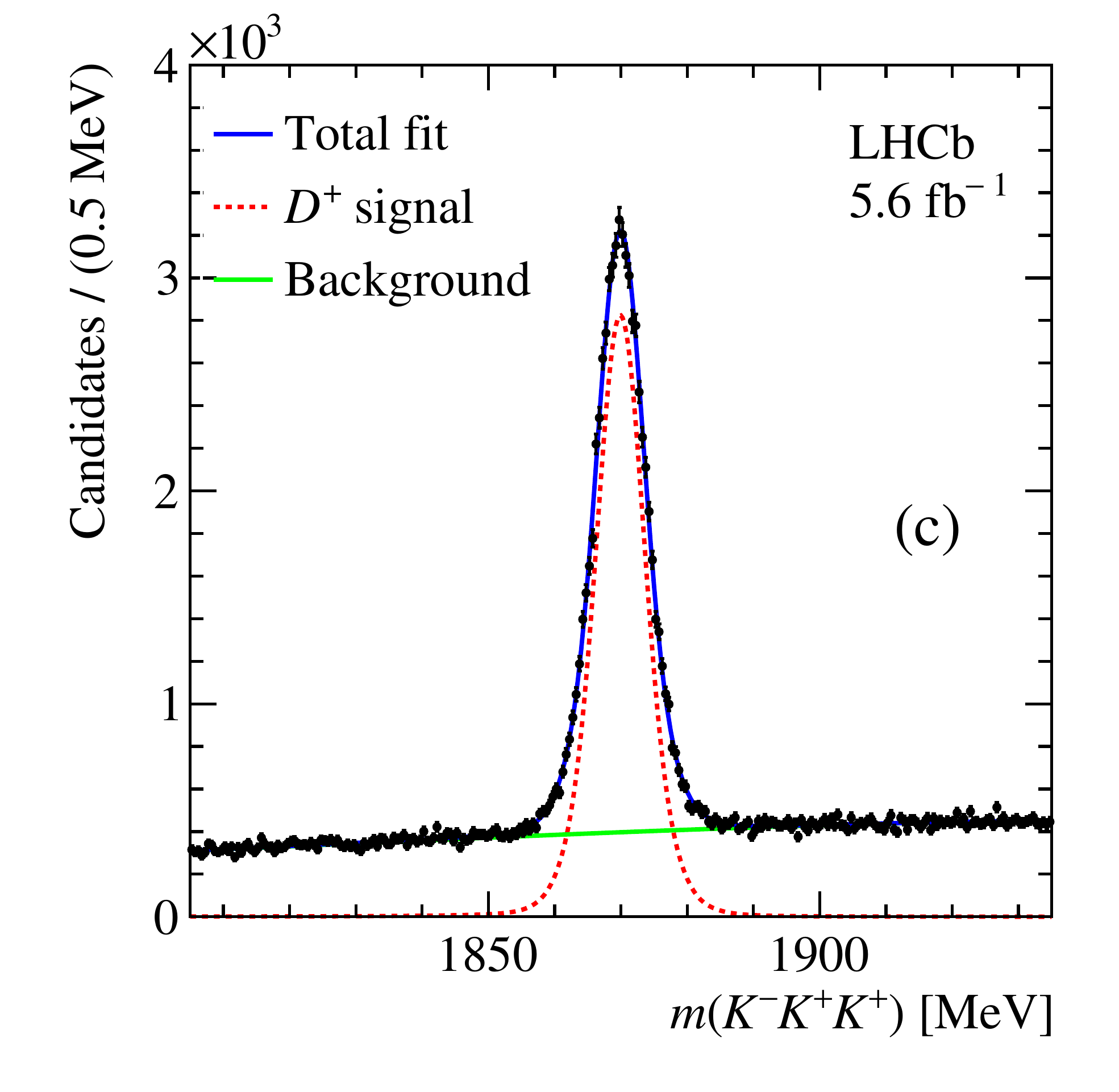}
    \includegraphics[width=0.45\linewidth]{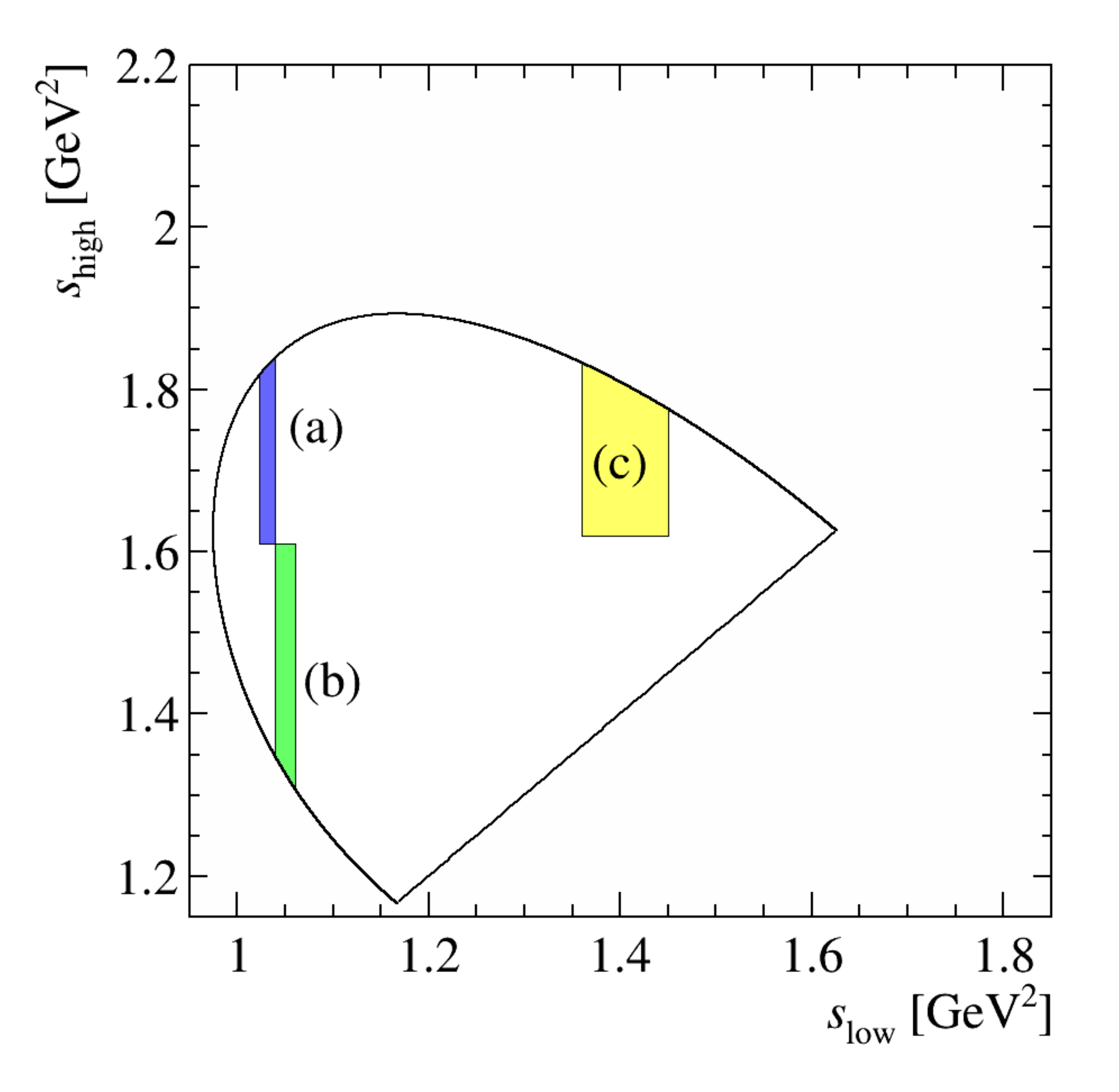}
    \vspace*{-0.5cm}
  \end{center}
  \caption{
    \small Mass distributions for the $\Dp$ candidates in three representative Dalitz plot bins (a,b,c), defined in the lower right subfigure.}
  \label{fig:dkkk-binfit}
\end{figure}

The binning scheme utilised in this analysis consists of 21 bins for each decay channel,\footnote{Sensitivity studies with pseudoexperiments show that for this dataset size the optimal number of bins is about 20: fewer bins may result in loss of sensitivity by integrating potential local effects with opposite signs, while more bins lead to a loss of statistical power.}  chosen such that  the number of signal candidates varies within roughly a factor of two among all bins.  The division of the bins around $s_{\rm low}\sim 1\gevgev$ is defined to enhance the sensitivity to \CP asymmetries that may change sign when crossing the $\phi(1020)$ resonance mass or the node of its angular distribution,  for instance due to S- and P-wave interference~\cite{LHCb-PAPER-2014-044,LHCb-PAPER-2019-018}. 
 The binning maps can be seen in Fig.~\ref{fig:21bingrids}. An alternative binning with 50 bins is used for cross-checks.
For the $\Km\Kp\Kp$ invariant-mass fits in each Dalitz-plot bin, the signal and background models are those described in Sect.~\ref{sec:data}, where the signal shape parameters are fixed to values obtained from simulation in each bin.
Fits to the $\Km\Kp\Kp$ and $\Kp\Km\Km$ mass spectra in each bin are performed independently.
Figures~\ref{fig:dskkk-binfit}~and~\ref{fig:dkkk-binfit} show examples of the mass fits for three representative bins.


\begin{figure}[htpb!]

\centering
  \includegraphics[width=0.49\textwidth]{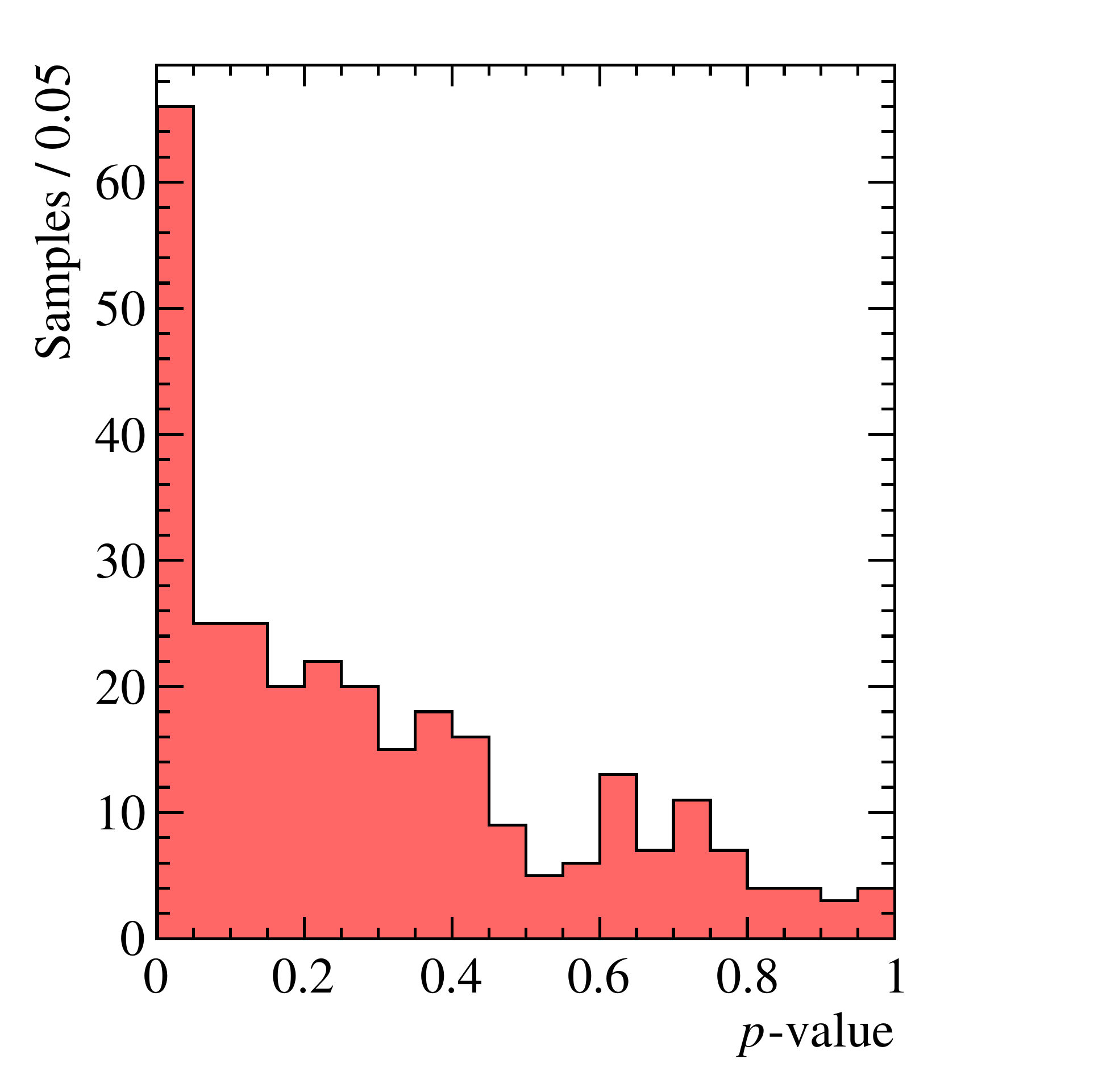}
  \includegraphics[width=0.49\textwidth]{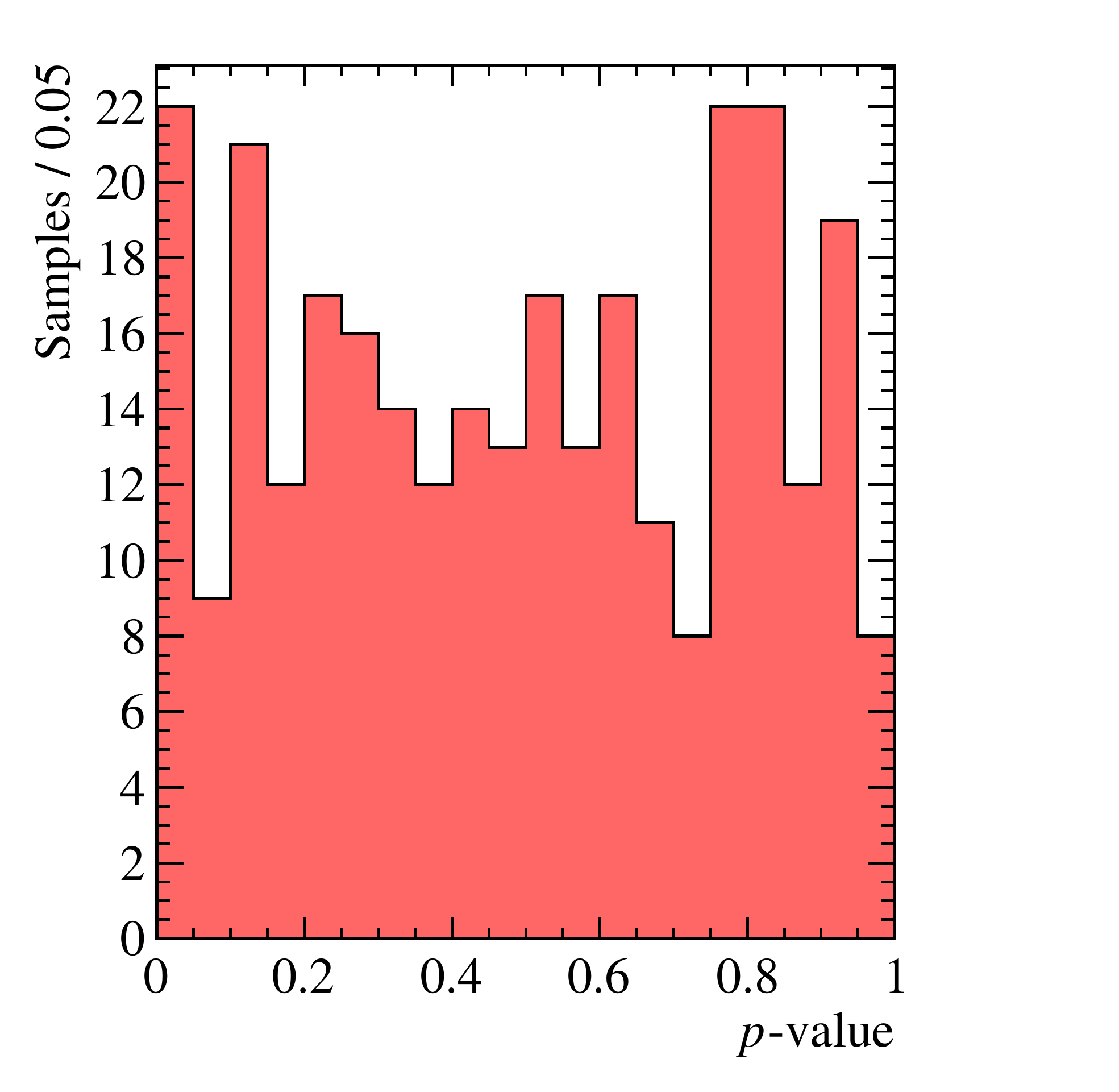}
 \caption{Distributions of the \textit{p}-values for (left) the original Miranda method and for (right) the fit-per-bin method using 300 \DsptoKKK  pseudoexperiments.}
\label{fig:dskkk-scp-null-pval}
 \end{figure}

Studies are performed to validate the  fit-per-bin method. Pseudoexperiments are generated with \Km\Kp\Kp  mass distributions following those obtained in data for \Dsp and \Dsm decays, for the 21 Dalitz-plot bins. The total number of events is varied according to a Poisson distribution based on the yields obtained from fitting the samples in each bin without charge separation. The number of candidates, for both signal and background,  is scaled by the relative fraction of \Dsp and \Dsm candidates in the full data sample, in order to introduce an integrated asymmetry. With this test, it is possible to compare the response of the original Miranda method with the  fit-per-bin method. The distributions of the \textit{p}-values for the \DsptoKKK decay following the two methodologies can be seen in Fig.~\ref{fig:dskkk-scp-null-pval}. For the current dataset size and level of background, the  original Miranda method introduces a bias towards lower $p$-values whereas the fit-per-bin method leads to a uniform distribution  as expected for the null-hypothesis.  The fit-per-bin distribution of \textit{p}-values is also seen to be uniform for \DptoKKK pseudoexperiments.

A cross-check of the fit-per-bin method is also performed using data, where the final  samples are divided randomly in two parts according to the overall proportions of \Dspp and \Dsmm but with no charge separation. The method is applied for both channels using  the baseline 21-bin scheme and also the alternative 50-bin scheme. The resulting \textit{p}-values lie between $30\%$ and $91\%$, compatible with the null-hypothesis.

The approximate sensitivity of the current data samples to \CP-violating effects is studied using simulated pseudoexperiments with similar resonant structure~\cite{LHCb-PAPER-2018-039} and number of signal and background candidates as in data. In these pseudoexperiments, phases or relative magnitudes
 are set to different values in the key resonant states, such as $\phi(1020)\Kp$ and $f_0(980)\Kp$, between the \Dspp and \Dsmm amplitudes.  It is found that the $p$-value threshold of $3\times 10^{-7}$ can be achieved for a difference of $3^\circ$ to $7^\circ$ in phase  or $3\%$ to $7\%$ in relative magnitude depending on the resonant state for which these effects are introduced in \Dspp and \Dsmm decays.
These input differences translate to \CP asymmetries typically of the order of a few percent. For the CS mode \DsptoKKK effects of this order are in principle still much higher than the SM expectation~(see for instance Ref.~\cite{Cheng:2021yrn}).

\section{Sensitivity to nuisance asymmetries}

The  sensitivity to possible nuisance localised charge asymmetries induced by detection efficiency or production rate is assessed through studies using fully simulated samples, the $\Km\Kp\Kp$ background candidates in data,   and Cabibbo-favoured control samples. 

Using simulation samples, about 2.7  and 5.4 million \DsptoKKK and  \mbox{\DptoKKK} candidates, respectively,  are retained after applying the same selection requirements as those in data. Since there is no background and the samples were generated with a flat distribution in phase space, the original Miranda method is used with a 21 same-size binning scheme.  The $p$-values found are 37.7\% and 76.4\% for the  \DsptoKKK and \DptoKKK decays,  respectively, showing no induced asymmetries from the detector responses, according to simulation.

For the background study, there are in total about 3.4 and 2.5  million  candidates in the \Dsp and \Dp samples, respectively. The consistency with no charge asymmetry  is tested using the fit-per-bin method described in Sect.~\ref{sec:method}, where the signal yields and uncertainties are replaced with those of the  background from the fits per bin. The resulting \textit{p}-values are 46.0\% and 75.7\% for the \Dsp and \Dp backgrounds, respectively, which indicate no effect of local charge symmetries.

\begin{figure}[tb]
  \begin{center}
    \includegraphics[width=0.49\linewidth]{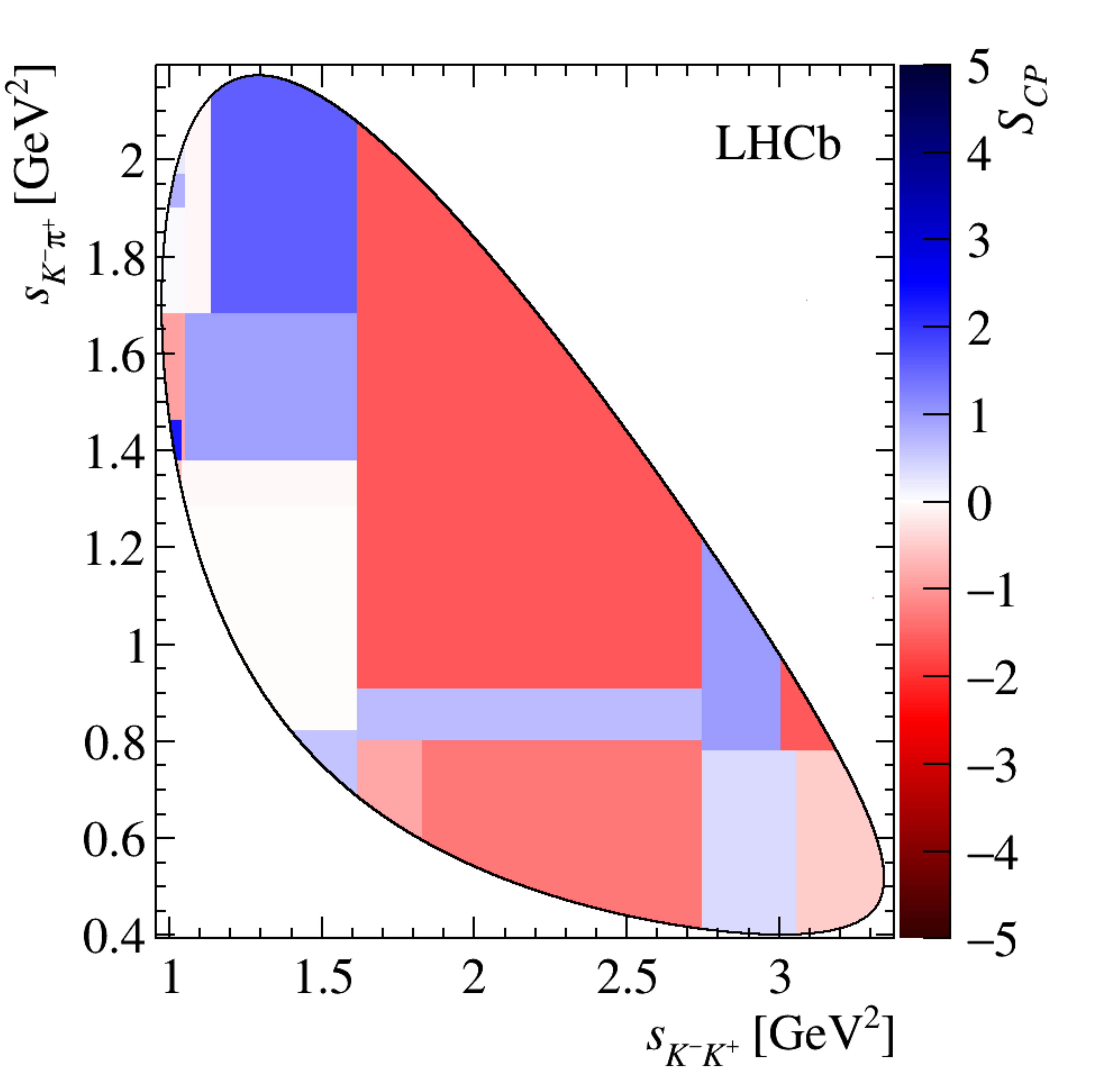}
    \includegraphics[width=0.49\linewidth]{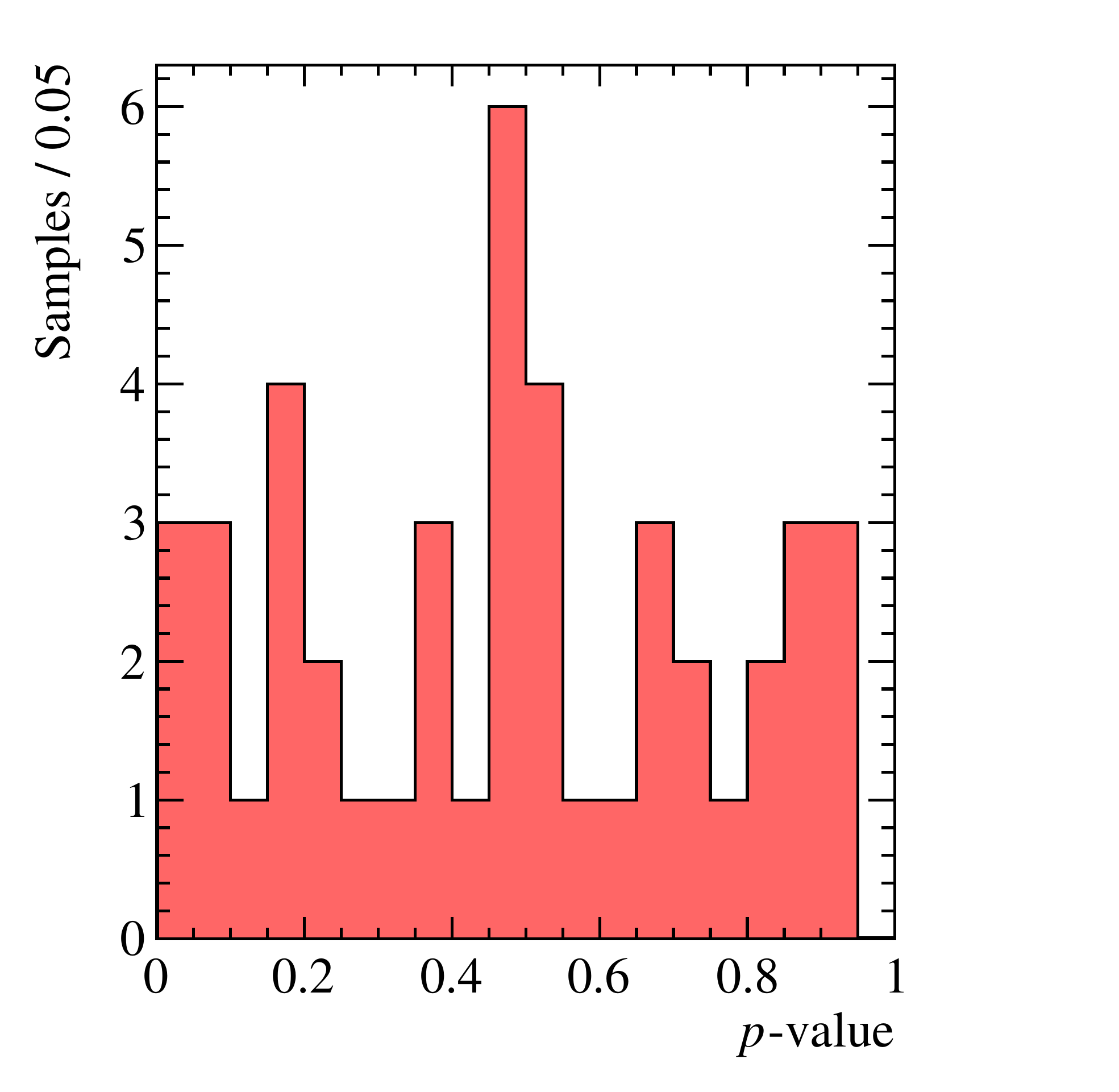}
    \vspace*{-0.5cm}
  \end{center}
  \caption{
    \small 
Left: Values of $\mathcal{S}_{\CP}$ for one subset of \DsptoKKP data with \DsptoKKK selection criteria. Right: Overall \textit{p}-value distribution for the 45 \DsptoKKP subsets using the scheme of  21 adaptive bins.
    }
  \label{fig:dskkp-control}
\end{figure}

\begin{figure}[tb]
  \begin{center}
  \includegraphics[width=0.49\linewidth]{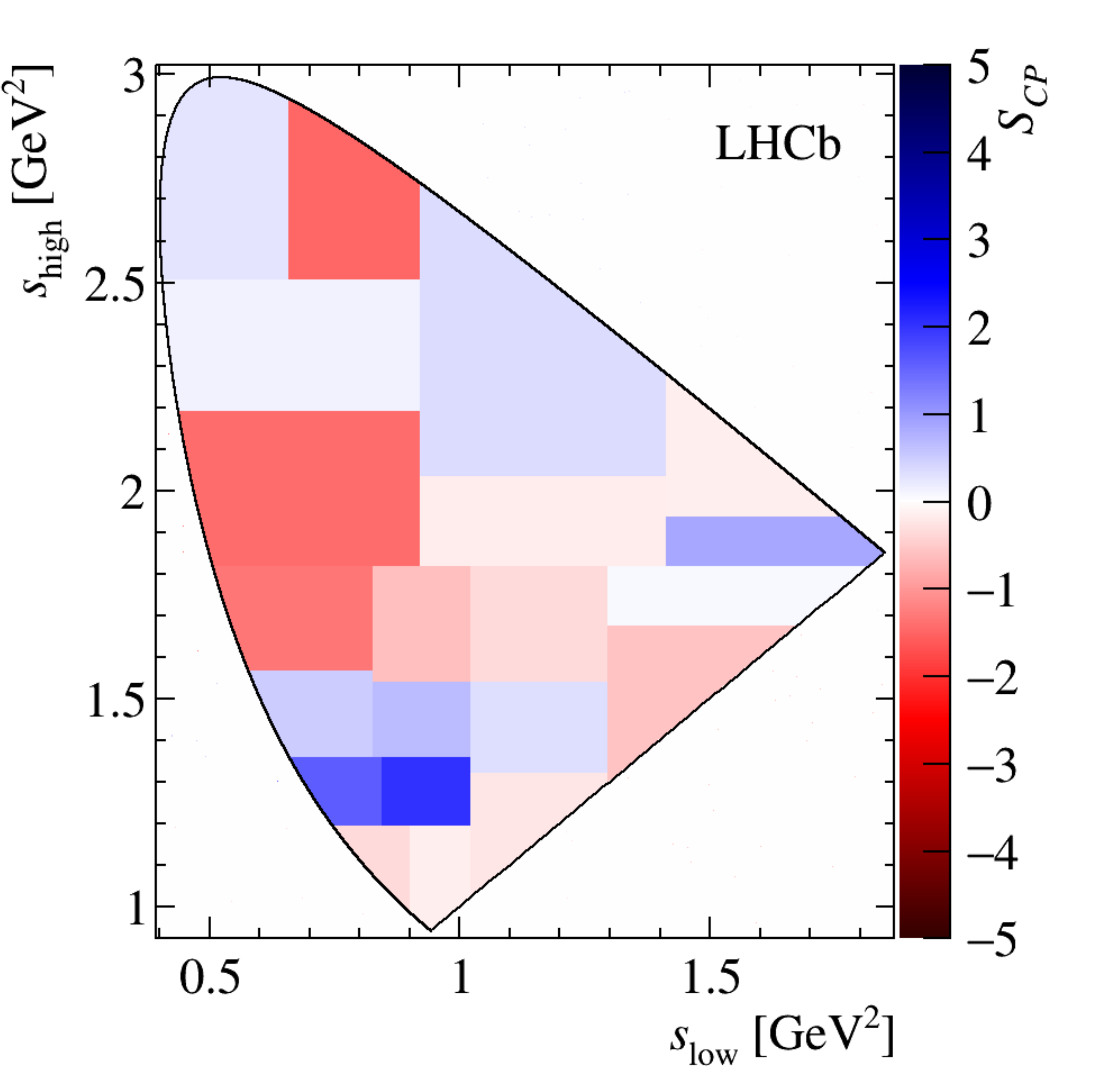}
    \includegraphics[width=0.49\linewidth]{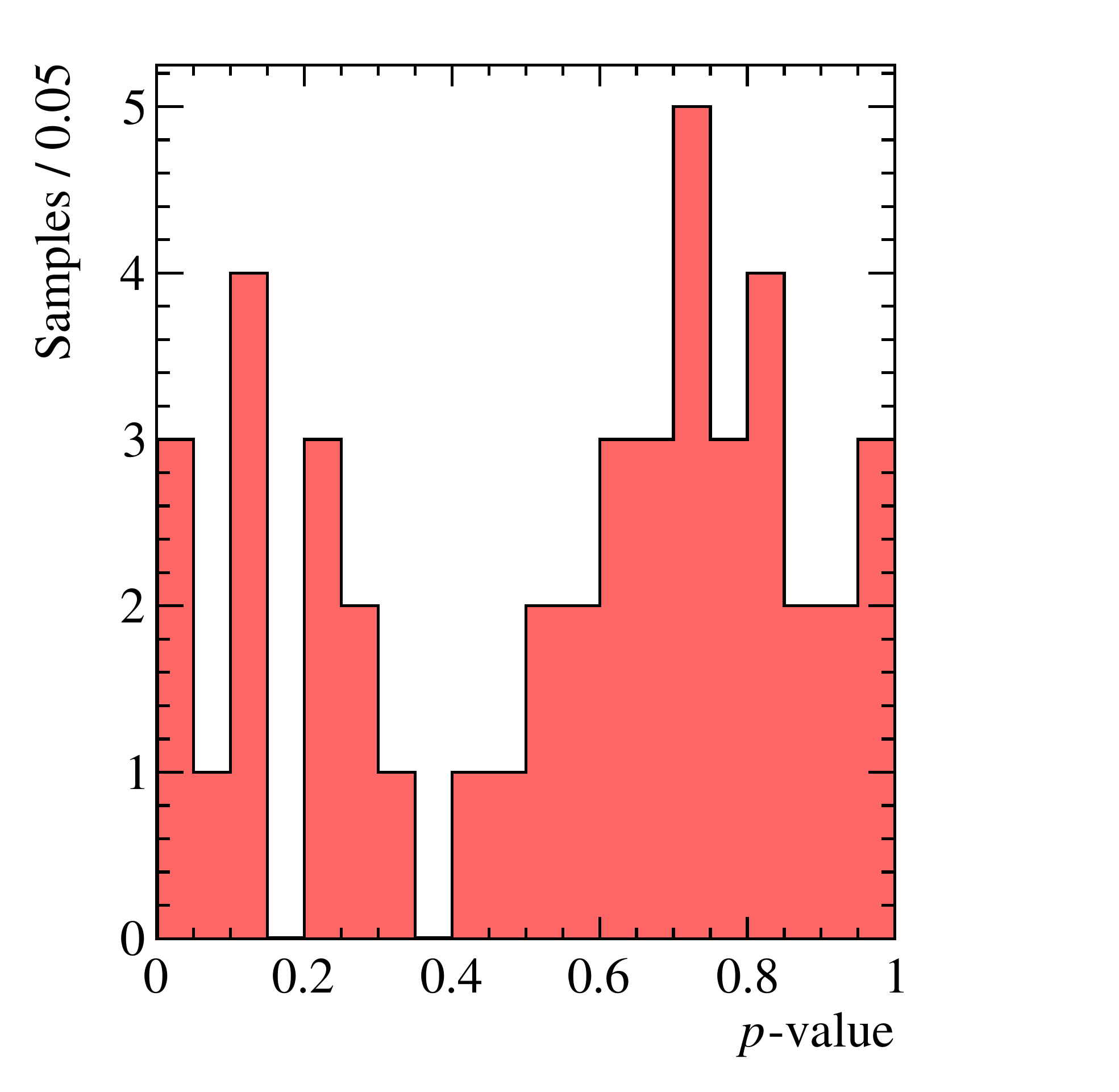}
    \vspace*{-0.5cm}
  \end{center}
  \caption{
    \small Left: Values of $\mathcal{S}_{\CP}$ distribution for one subset of \DptoKPP data with \DsptoKKK selection criteria. Right: Overall \textit{p}-value distribution for the 85 \DptoKPP subsets using the  scheme of  21 adaptive bins.
    }
  \label{fig:dkpp-control}
\end{figure}

The Cabibbo-favoured \DsptoKKP and \DptoKPP decays are used as control channels. These modes have much larger data samples than the analysed signal samples. To test whether local nuisance asymmetries are present in the data,  
the control samples are divided into multiple subsets of approximately five times the size of the signal channels. There are 45 samples of \DsptoKKP decays and 85 of \DptoKPP decays with over $90\%$ purity. Since the selection criteria are different for the \DsptoKKK and \DptoKKK samples, the control-channel samples are studied applying both sets of criteria (except the PID requirements for the pion candidates). To perform the search for local asymmetries, four binning schemes are tested: two uniform grids, of $5\times 5$ and $8\times 8$, and two adaptive binnings such that all bins have the same population, with 21 and 50 bins in total. Due to the high purity of the samples, the original Miranda technique is used, \ie simple counting to obtain the number of \Dsp and \Dp candidates, respectively,  in the $m(\Km\Kp\pip)$ region of 1950--1980\mev and in the $m(\Km\pip\pip)$ region of 1850--1890\mev. The resulting \textit{p}-value distributions for both control samples under the different selection criteria and binning schemes are found to be uniformly distributed, with no value incompatible with the null hypothesis. These results are thus consistent with the absence of local nuisance charge asymmetries. Figures~\ref{fig:dskkp-control} and~\ref{fig:dkpp-control} show examples of the $\mathcal{S}_{\CP}$ distribution for one control-mode subset and  the overall \textit{p}-value distribution for the adaptive 21-bin scheme  for the \DsptoKKP  and the \DptoKPP channels where the selection criteria of the \DsptoKKK channel are applied to both.

\section{Results}
\label{sec:results}

\begin{figure}[tb]
  \centering
  \includegraphics[width=.49\linewidth]{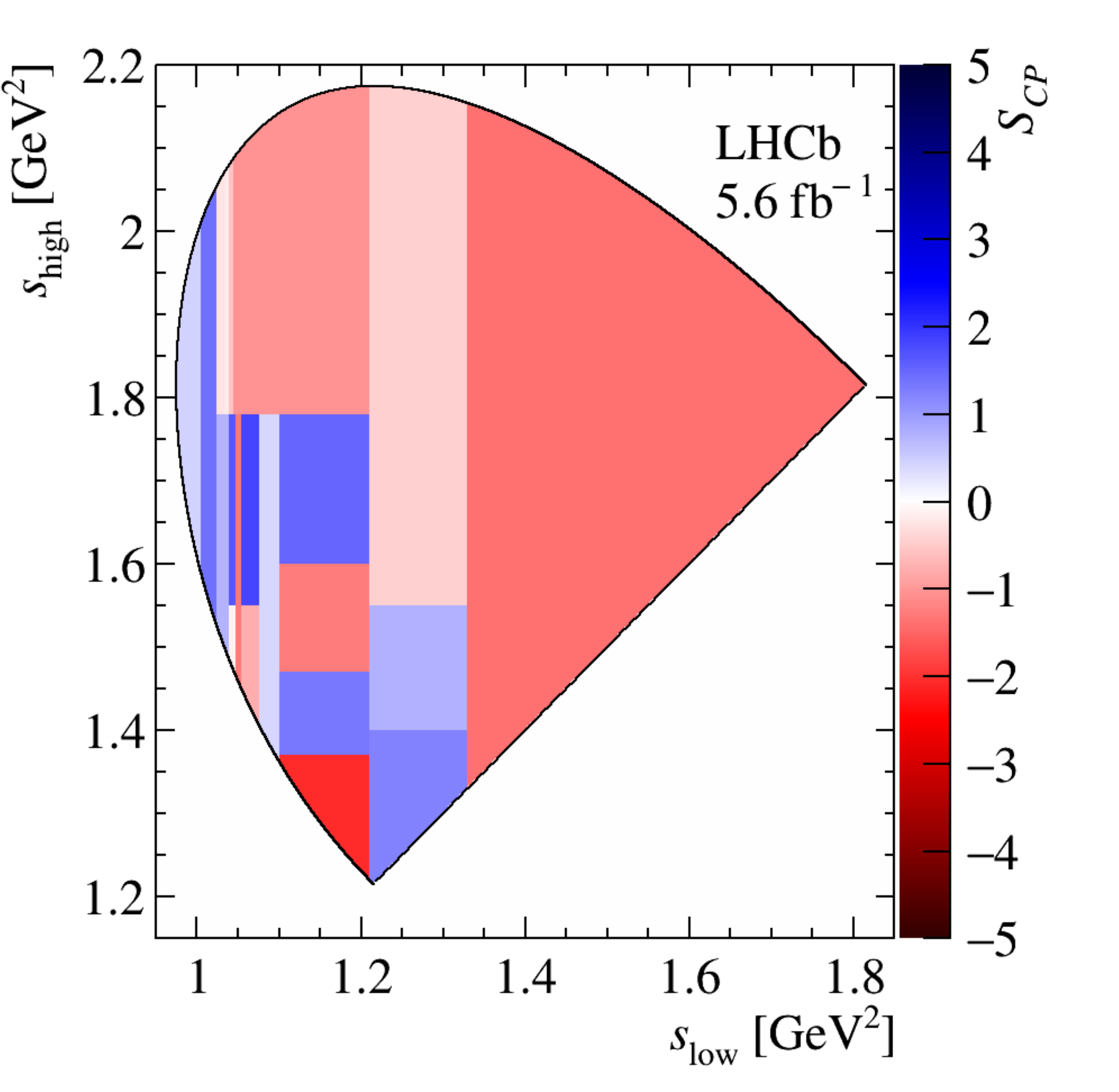}
    \includegraphics[width=.49\linewidth]{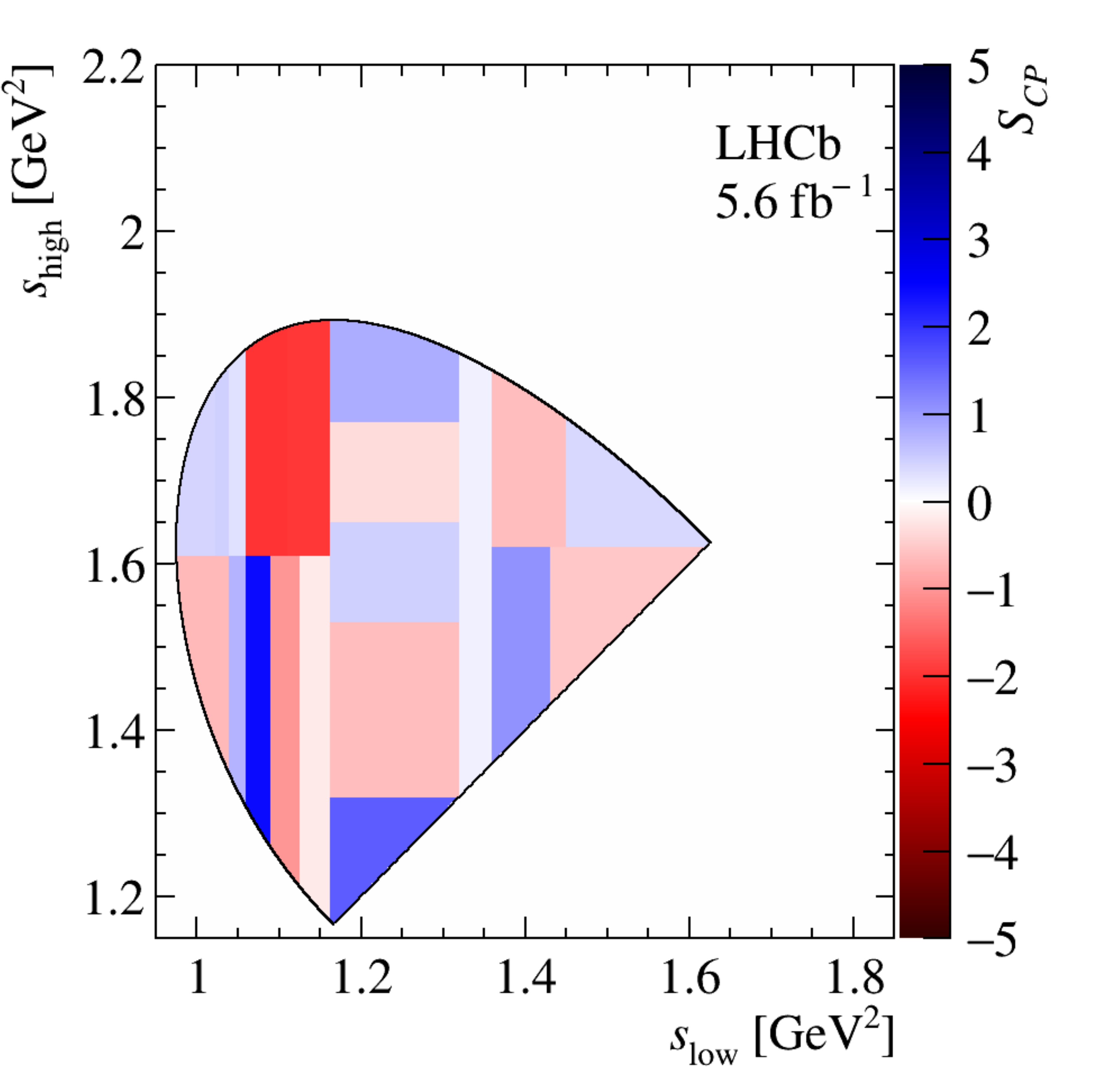}
 \caption{$\mathcal{S}_{\CP}$ values across the Dalitz plot for (left)  \DsptoKKK and (right) \mbox{\DptoKKK} signal candidates using 21 bins.}
 \label{fig:unblind-SCP}
 \end{figure}

Searches for CPV in the $\Dspp\to\Km\Kp\Kp$ decays are performed with the fit-per-bin strategy described in Sect.~\ref{sec:method}. 
The total signal yields for each charge  and the values of $\alpha$ for the \DsptoKKK and \DptoKKK channels are displayed in Table~\ref{tab:unblind-yields}.
The distributions of $\mathcal{S}_{\CP}$  for both decays  are shown in Fig.~\ref{fig:unblind-SCP}. The resulting \textit{p}-values are  13.3\% for the decay \DsptoKKK and 31.6\% for the decay \DptoKKK. The results are consistent with the hypothesis of no  localised \CP  violation in either channel.

\begin{table}[tb]
\centering
\caption{Dalitz-integrated $\Dspp$ and $\Dsmm$ signal yields and resulting values of $\alpha$.}
\begin{tabular}{c c c}\hline
  & \DsptoKKK & \DptoKKK \\ \hline
  $\sum_i N^i(D_{(s)}^+)$ & $(487.8 \pm 1.1)\times 10^3 $ & $(638.8\pm 1.1)\times 10^3$ \\
  $\sum_i N^i(D_{(s)}^-)$ & $(484.4 \pm 1.1)\times 10^3 $ & $(631.4 \pm 1.0)\times 10^3$ \\
  $\alpha$ & $1.007\pm0.003$ & $1.012\pm0.002$ 
\\ \hline
\end{tabular}
\label{tab:unblind-yields}
\end{table}

Potential systematic effects are probed through variations on the signal and background descriptions as well as on the binning scheme, and obtaining the resulting \textit{p}-values.
The signal shape is changed to two Gaussians plus two CB functions,  the background is changed to a third-order Chebyshev polynomial, and combinations of these changes are also tried. Additionally, instead of independent fits, simultaneous fits where the background shape parameters are shared among $\Dspp$ and $\Dsmm$ samples are also tested.  
Under these variations, the \textit{p}-values for the \DsptoKKK decay range from  1.7\% to 20.2\%, while for \DptoKKK they range from 3.2\% to 48.5\%.


\begin{figure}[tb]
  \centering
  \includegraphics[width=.49\linewidth]{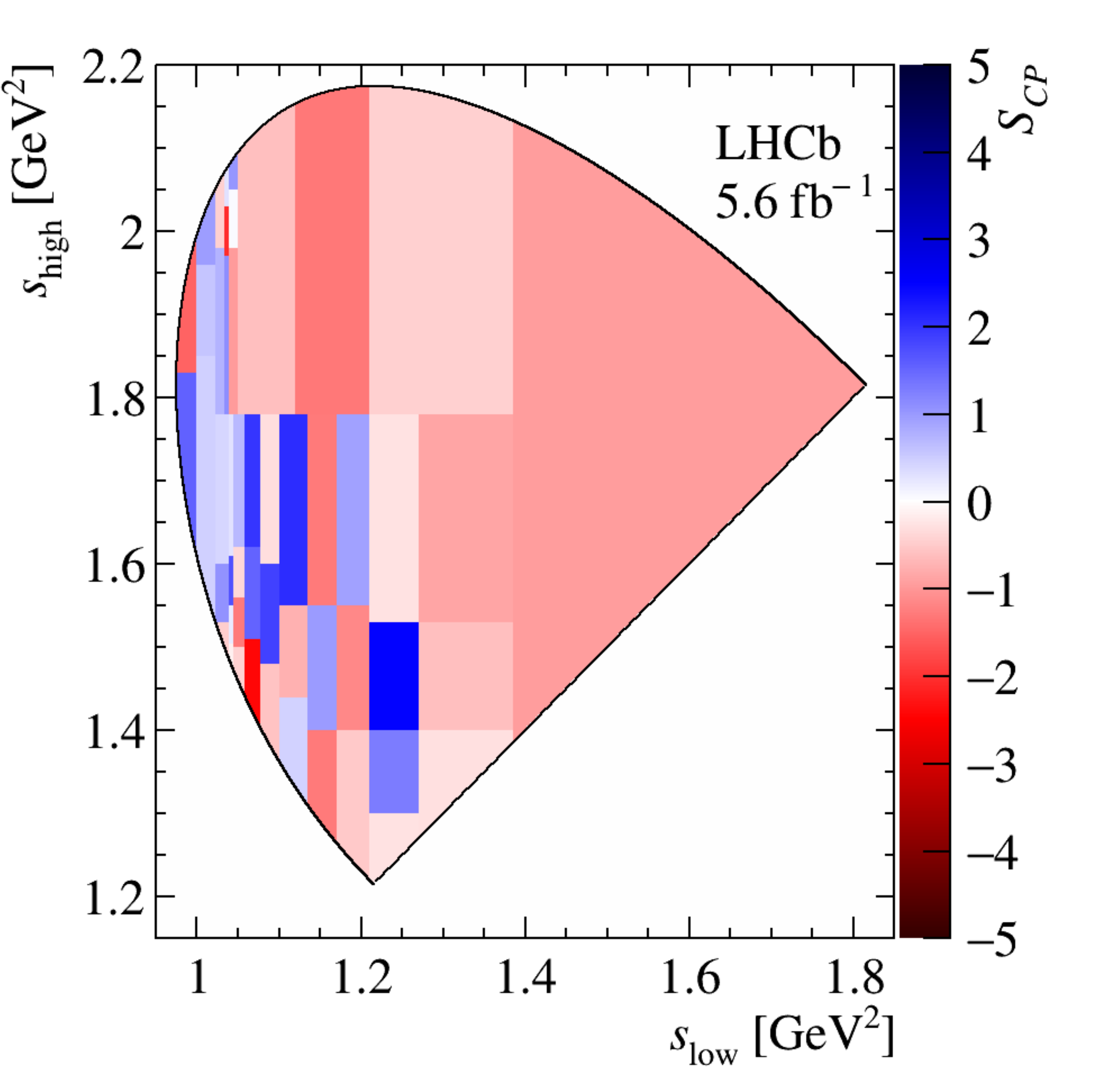}
    \includegraphics[width=.49\linewidth]{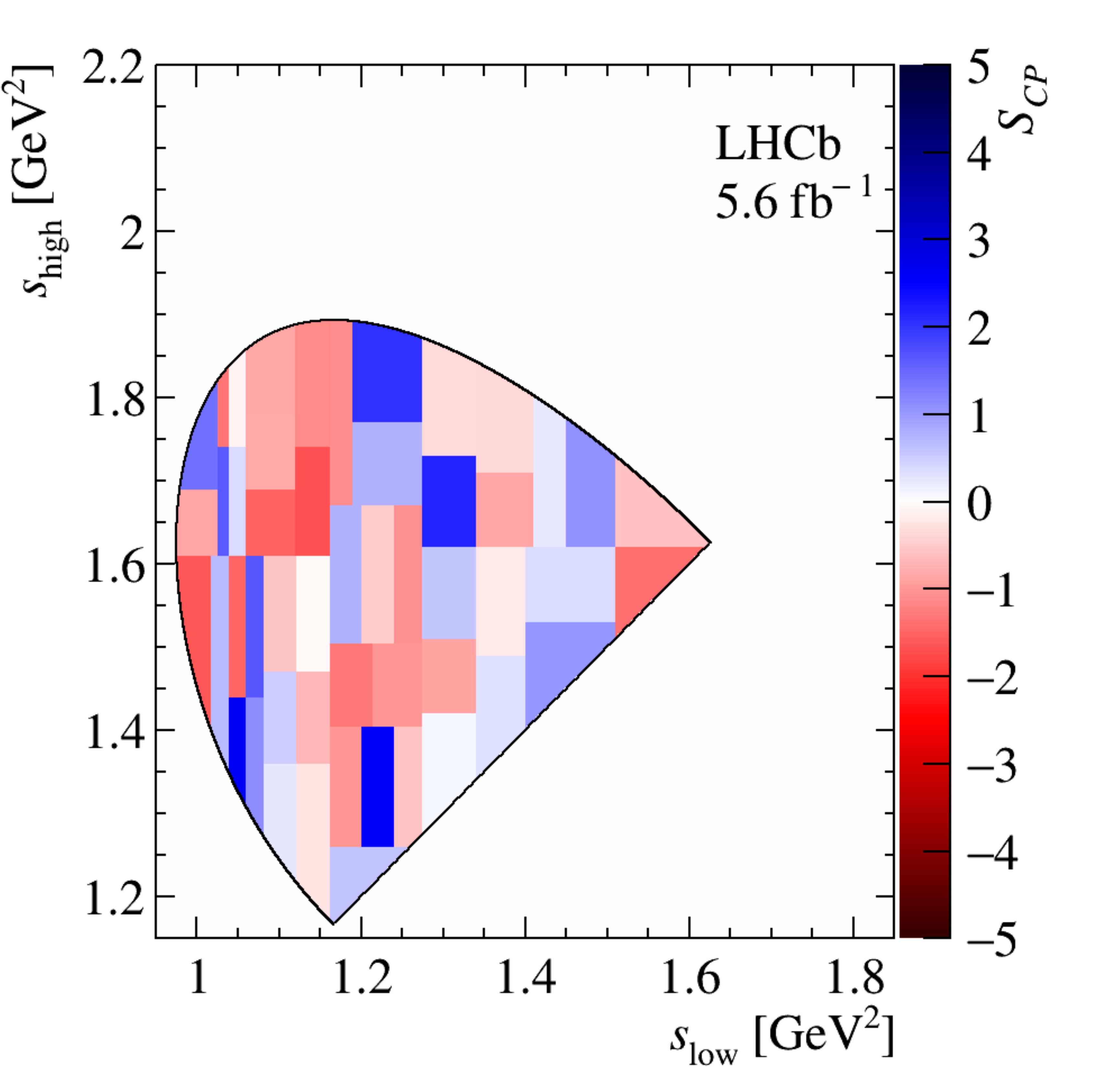}
 \caption{$\mathcal{S}_{\CP}$ values across the Dalitz plot for (left)  \DsptoKKK and (right) {\mbox\DptoKKK} signal candidates for the alternative binning scheme with 50 bins.}
 \label{fig:unblind-SCP-50}
 \end{figure}

An alternative binning scheme is tested, with 50 bins, using the baseline models for signal and background. A finer binning may allow a higher sensitivity in localised regions, though it could lead to an overall loss of statistical sensitivity. The $\mathcal{S}_{\CP}$ distributions for this alternative binning are shown in Fig.~\ref{fig:unblind-SCP-50}. The resulting \textit{p}-values are 10.1\% and 8.2\% for the \DsptoKKK and \DptoKKK channels, respectively. 

Altogether the results consistently indicate no evidence for \CP violation in the current samples for \DsptoKKK and \DptoKKK decays.

 \section{Conclusions}
\label{sec:conclusion}

 This paper presents a search for \CP violation in the Cabibbo-suppressed decay \mbox{\DsptoKKK} and in the doubly Cabibbo-suppressed decay \DptoKKK  using $pp$ collison data at $\sqrt{s}= 13\tev$ and corresponding to an integrated luminosity of 5.6\invfb. The study is performed with samples containing about $0.97$ ($1.27$) million $\Dsp(\Dp)\to \Km\Kp\Kp$ decays. A novel technique is used: the fit-per-bin method, which is a variation of the  original Miranda technique, where the value of $\mathcal{S}_\CP$ is calculated in each Dalitz plot bin using the \Dspp and \Dsmm signal yields estimated from $\Km\Kp\Kp$ mass-invariant fits.  The  fit-per-bin method is found to be robust for cases with very large signal samples and significant background. 

 The results are given as  \textit{p}-values with respect to the null-hypothesis of \CP conservation and are found to be 13.3\% for the  \DsptoKKK channel and 31.6\% for the \mbox{\DptoKKK} channel. No evidence for \CP violation is found. Cross-checks by varying the invariant-mass fit models and binning schemes give consistent results. 

 This is the first search for \CP violation in the Cabibbo-suppressed channel \mbox{\DsptoKKK} and in the doubly Cabibbo-suppressed channel \DptoKKK. The work presented here contributes to a wider effort of comprehensive studies of  \CP violation in charm decays.

\section*{Acknowledgements}
%
%
\noindent We express our gratitude to our colleagues in the CERN
accelerator departments for the excellent performance of the LHC. We
thank the technical and administrative staff at the LHCb
institutes.
We acknowledge support from CERN and from the national agencies:
CAPES, CNPq, FAPERJ and FINEP (Brazil); 
MOST and NSFC (China); 
CNRS/IN2P3 (France); 
BMBF, DFG and MPG (Germany); 
INFN (Italy); 
NWO (Netherlands); 
MNiSW and NCN (Poland); 
MEN/IFA (Romania); 
MICINN (Spain); 
SNSF and SER (Switzerland); 
NASU (Ukraine); 
STFC (United Kingdom); 
DOE NP and NSF (USA).
We acknowledge the computing resources that are provided by CERN, IN2P3
(France), KIT and DESY (Germany), INFN (Italy), SURF (Netherlands),
PIC (Spain), GridPP (United Kingdom), 
CSCS (Switzerland), IFIN-HH (Romania), CBPF (Brazil),
Polish WLCG  (Poland) and NERSC (USA).
We are indebted to the communities behind the multiple open-source
software packages on which we depend.
Individual groups or members have received support from
ARC and ARDC (Australia);
Minciencias (Colombia);
AvH Foundation (Germany);
EPLANET, Marie Sk\l{}odowska-Curie Actions and ERC (European Union);
A*MIDEX, ANR, IPhU and Labex P2IO, and R\'{e}gion Auvergne-Rh\^{o}ne-Alpes (France);
Key Research Program of Frontier Sciences of CAS, CAS PIFI, CAS CCEPP, 
Fundamental Research Funds for the Central Universities, 
and Sci. \& Tech. Program of Guangzhou (China);
GVA, XuntaGal, GENCAT and Prog.~Atracci\'on Talento, CM (Spain);
SRC (Sweden);
the Leverhulme Trust, the Royal Society
 and UKRI (United Kingdom).

\clearpage

\addcontentsline{toc}{section}{References}
\bibliographystyle{LHCb}
\bibliography{main,standard,LHCb-PAPER,LHCb-CONF,LHCb-DP,LHCb-TDR,charm_theory}

\newpage
\centerline
{\large\bf LHCb collaboration}
\begin
{flushleft}
\small
R.~Aaij$^{32}$\lhcborcid{0000-0003-0533-1952},
A.S.W.~Abdelmotteleb$^{50}$\lhcborcid{0000-0001-7905-0542},
C.~Abellan~Beteta$^{44}$,
F.~Abudin{\'e}n$^{50}$\lhcborcid{0000-0002-6737-3528},
T.~Ackernley$^{54}$\lhcborcid{0000-0002-5951-3498},
B.~Adeva$^{40}$\lhcborcid{0000-0001-9756-3712},
M.~Adinolfi$^{48}$\lhcborcid{0000-0002-1326-1264},
P.~Adlarson$^{77}$\lhcborcid{0000-0001-6280-3851},
H.~Afsharnia$^{9}$,
C.~Agapopoulou$^{13}$\lhcborcid{0000-0002-2368-0147},
C.A.~Aidala$^{78}$\lhcborcid{0000-0001-9540-4988}
% ...

{\footnotesize \it

% ...
$^{9}$Universit{\'e} Clermont Auvergne, CNRS/IN2P3, LPC, Clermont-Ferrand, France\\
% ...
$^{13}$LPNHE, Sorbonne Universit{\'e}, Paris Diderot Sorbonne Paris Cit{\'e}, CNRS/IN2P3, Paris, France\\
% ...
$^{32}$Nikhef National Institute for Subatomic Physics, Amsterdam, Netherlands\\
% ...
$^{40}$Instituto Galego de F{\'\i}sica de Altas Enerx{\'\i}as (IGFAE), Universidade de Santiago de Compostela, Santiago de Compostela, Spain\\
% ...
$^{44}$Physik-Institut, Universit{\"a}t Z{\"u}rich, Z{\"u}rich, Switzerland\\
% ...
$^{48}$H.H. Wills Physics Laboratory, University of Bristol, Bristol, United Kingdom\\
% ...
$^{50}$Department of Physics, University of Warwick, Coventry, United Kingdom\\
% ...
$^{53}$School of Physics and Astronomy, University of Glasgow, Glasgow, United Kingdom\\
$^{54}$Oliver Lodge Laboratory, University of Liverpool, Liverpool, United Kingdom\\
% ...
$^{62}$Syracuse University, Syracuse, NY, United States\\
% ...
$^{77}$Department of Physics and Astronomy, Uppsala University, Uppsala, Sweden, associated to $^{53}$\\
$^{78}$University of Michigan, Ann Arbor, MI, United States, associated to $^{62}$\\
\bigskip
% ...
}
\end{flushleft}
\end{document}